\documentclass[a4paper,onecolumn,11pt]{quantumarticle}
\pdfoutput=1

\usepackage{amsmath,amssymb,mathtools}
\usepackage{amsthm}
\usepackage{booktabs}
\usepackage{graphicx}
\usepackage{subcaption}
\usepackage[numbers,sort&compress]{natbib}
\usepackage[acronym,nomain]{glossaries}
\usepackage{hyperref}

\newtheorem{theorem}{Theorem}

\newtheorem{lemma}{Lemma}
\newtheorem{corollary}{Corollary}

\newacronym{QPD}{QPD}{quasiprobability decomposition}

\title{Optimal Two-Qubit Gate-Cutting Cost and Measures of Nonlocality}

\author{Michael Hart}
\email{mhart727@qub.ac.uk}

\begin{document}

\maketitle

\begin{abstract}
Circuit cutting \cite{Peng_2020} enables quantum computations to be decomposed into smaller subcircuits at the cost of additional sampling overhead. Gate-cutting \cite{Mitarai_2019} provides one such approach by replacing nonlocal gates with quasiprobability decompositions of local operations. Starting from the known optimal two-qubit gate-cutting formula, we derive complete sharp lower and upper envelopes relating the \textit{quasiprobability extent} \cite{piveteau2025circuitcuttingclassicalinformation} $\gamma$ to five established descriptors: entangling power, gate typicality, operator entanglement, Schmidt strength, and maximum product-input concurrence. We recast $\gamma$ as a R\'enyi-$1/2$ functional of the operator-Schmidt spectrum and show how the physical constraints on two-qubit spectra sharpen generic entropy bounds and select a small set of recurring extremal Cartan families. None of the five descriptors generically determines $\gamma$ alone, but each imposes exact constraints, revealing how substantially the optimal gate-cutting cost can vary between gates with similar nonlocal characteristics.
\end{abstract}

% ================================================================
\section{Introduction}
\label{sec:introduction}
% ================================================================

Circuit cutting evaluates quantum computations that are too large or
poorly connected to execute directly by partitioning them into smaller
subcircuits and reconstructing global quantities from separate
experiments \cite{Peng_2020}. Quasiprobability-based schemes incur
sampling overhead; in gate cutting, a nonlocal gate is replaced by a
\gls{QPD} of locally implementable operations
\cite{Mitarai_2019,Mitarai_2021}. Schmitt, Piveteau and Sutter give the
exact optimal value of $\gamma$ for arbitrary two-qubit unitaries and
extend it to joint cutting of multiple two-qubit gates
\cite{Schmitt_2025}. Harrow and Lowe connect the related product extent
to optimal space-like cutting for cases including all two-qubit gates
\cite{Harrow_2025}, while Piveteau's thesis develops the
quasiprobability-extent framework for nonlocal states and channels,
including state-conversion lower bounds
\cite{https://doi.org/10.3929/ethz-b-000727956}. Following
Ref.~\cite{piveteau2025circuitcuttingclassicalinformation}, we call the
optimal gate-cutting quantity considered here the
\emph{quasiprobability extent}, $\gamma$; below, simply the
\gls{QPD} extent.

We ask instead how strongly the intrinsic nonlocal structure of a
two-qubit gate constrains its optimal \gls{QPD} extent. Two-qubit
local-equivalence classes admit a three-parameter Cartan description
\cite{PhysRevA.67.042313}, but are also summarised by scalar
descriptors of different aspects of nonlocal action. We consider
entangling power \cite{Zanardi_2000}, gate typicality
\cite{Jonnadula_2020}, operator entanglement
\cite{Jonnadula_2020,Musz_2013}, Schmidt strength
\cite{Nielsen_2003}, and maximum product-input concurrence
\cite{Kraus_2001,Chefles_2005}. Rather than deriving a new cutting
formula, we determine the exact minimum and maximum extent at every
fixed value of each descriptor.

Specifically, we make the following contributions:
\begin{enumerate}
    \renewcommand{\labelenumi}{(\roman{enumi})}
    \item Starting from the optimal two-qubit cutting formula of
    Ref.~\cite{Schmitt_2025}, we recast the \gls{QPD} extent as a
    R\'enyi-$1/2$ functional of the normalised operator-Schmidt
    spectrum. Using the unistochastic characterisation of physical
    two-qubit spectra \cite{Musz_2013}, we rewrite the physicality
    condition in an ordered form convenient for the extremal analysis.
    The resulting condition also recovers the known exclusion of
    operator-Schmidt rank three
    \cite{Dur_2002,Tyson_2003}.

    \item We derive complete sharp lower and upper envelopes relating
    the \gls{QPD} extent to all five descriptors. These include the
    complete piecewise gate-typicality envelopes and physically
    sharpened operator-entanglement and Schmidt-strength bounds that are
    stricter than the corresponding generic probability-simplex entropy
    bounds.

    \item We identify the recurring canonical Cartan families that
    realise the sharp boundaries. For maximum concurrence, the general
    state-conversion lower bound of Ref.~\cite{https://doi.org/10.3929/ethz-b-000727956}
    yields the lower inequality; we complete the fixed-$C_{\max}$
    extremal characterisation by proving that both sharp boundaries are
    realised for every $C_{\max}$. The perfect-entangler range follows
    as a direct consequence.
\end{enumerate}

Section~\ref{sec:setup} introduces the Cartan and operator-Schmidt
framework and the five descriptors, and
Section~\ref{sec:canonical-families} collects the canonical gate
families. Sections~\ref{sec:gate-action}--\ref{sec:entangling-capacity}
derive the five pairs of sharp envelopes; Section~\ref{sec:discussion}
compares them, and Section~\ref{sec:conclusion} concludes. Detailed
proofs are collected in
Appendices~\ref{app:canonical-families}--\ref{app:cmax-perfect-entanglers}.

% ================================================================

\section{Two-Qubit Gates and Optimal Cutting Cost}
\label{sec:setup}
% ================================================================

Up to local single-qubit operations, any two-qubit unitary is specified
by three Cartan parameters, which also determine its operator-Schmidt
spectrum and hence the optimal \gls{QPD} extent. We first fix this
notation and then define the five nonlocality descriptors used below.

% ----------------------------------------------------------------
\subsection{Cartan representation}
\label{subsec:cartan}
% ----------------------------------------------------------------

Any two-qubit unitary $U$ admits a Cartan, or KAK, decomposition
\cite{PhysRevA.67.042313,Schmitt_2025}
\begin{equation}
    U
    =
    (V_1\otimes V_2)\,
    U_{\mathrm d}\,
    (V_3\otimes V_4),
    \label{eq:kak}
\end{equation}
where the $V_j$ are single-qubit unitaries and
\begin{equation}
    U_{\mathrm d}
    =
    \exp\!\left[
        \mathrm{i}\left(
        \theta_1 X\otimes X
        +\theta_2 Y\otimes Y
        +\theta_3 Z\otimes Z
        \right)
    \right].
    \label{eq:cartan-unitary}
\end{equation}
We use the signed Weyl-chamber convention of Schmitt, Piveteau and
Sutter \cite{Schmitt_2025},
\begin{equation}
    |\theta_3|
    \leq
    \theta_2
    \leq
    \theta_1
    \leq
    \frac{\pi}{4}.
    \label{eq:weyl-chamber}
\end{equation}
The local pre- and post-rotations do not affect the optimal two-qubit
\gls{QPD} extent, so it is sufficient to cut $U_{\mathrm d}$
\cite{Schmitt_2025}. The \gls{QPD} extent and the five quantities
introduced below are invariant under $\theta_3\mapsto-\theta_3$; this
follows directly from Eqs.~\eqref{eq:schmidt-probabilities-cartan},
\eqref{eq:ep-cartan}, \eqref{eq:gt-cartan}, and
\eqref{eq:cmax-cartan}. For analyses involving these quantities, we
may therefore restrict to the non-negative tetrahedron,
\begin{equation}
    0\leq\theta_3\leq\theta_2\leq\theta_1\leq\frac{\pi}{4}.
    \label{eq:positive-weyl}
\end{equation}

\begin{figure}[t]
    \centering
    \includegraphics[width=0.7\linewidth]{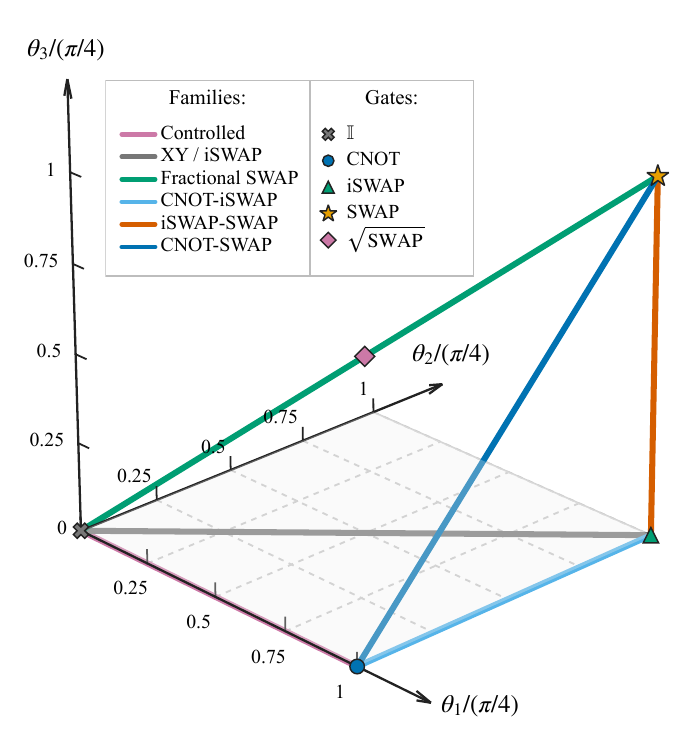}
    \caption{Non-negative Cartan tetrahedron of
    Eq.~\eqref{eq:positive-weyl}, with each Cartan coordinate
    normalised by $\pi/4$. The marked points show the identity, controlled-NOT (CNOT),
    iSWAP, SWAP and $\sqrt{\mathrm{SWAP}}$ classes, while the coloured
    edges identify the six canonical one-parameter families used
    throughout the extremal analysis.}
    \label{fig:cartan-tetrahedron}
\end{figure}

The resulting representative tetrahedron and the canonical gate
families used throughout the extremal analysis are shown in
Figure~\ref{fig:cartan-tetrahedron}.

% ----------------------------------------------------------------
\subsection{Operator-Schmidt representation and \texorpdfstring{\gls{QPD}}{QPD} extent}
\label{subsec:schmidt-gamma}
% ----------------------------------------------------------------

The three generators in Eq.~\eqref{eq:cartan-unitary} commute.
Expanding the interaction unitary in the corresponding Pauli-product
basis gives the representation used in Ref.~\cite{Schmitt_2025},
\begin{equation}
    U_{\mathrm d}
    =
    \sum_{j=0}^{3}
    u_j\,\sigma_j\otimes\sigma_j,
    \qquad
    \sigma_0=\mathbb I,\;
    \sigma_1=X,\;
    \sigma_2=Y,\;
    \sigma_3=Z,
    \label{eq:pauli-schmidt}
\end{equation}
where direct expansion of Eq.~\eqref{eq:cartan-unitary} gives
\begin{align}
    u_0 &=
    c_1c_2c_3
    + \mathrm{i} s_1s_2s_3,
    \nonumber\\
    u_1 &=
    c_1s_2s_3
    + \mathrm{i} s_1c_2c_3,
    \nonumber\\
    u_2 &=
    s_1c_2s_3
    + \mathrm{i} c_1s_2c_3,
    \nonumber\\
    u_3 &=
    s_1s_2c_3
    + \mathrm{i} c_1c_2s_3,
    \label{eq:kak-coefficients}
\end{align}
with $c_j=\cos\theta_j$ and $s_j=\sin\theta_j$.

After absorbing the phase of each $u_j$ into one local factor, the
Hilbert--Schmidt-orthonormal basis
$\{\sigma_j/\sqrt{2}\}_{j=0}^{3}$ gives operator-Schmidt singular
values $2|u_j|$. Following the normalised operator-Schmidt-vector
convention of Ref.~\cite{Musz_2013}, we therefore define
\begin{equation}
    p_j = |u_j|^2,
    \qquad
    \sum_{j=0}^{3}p_j=1,
    \label{eq:schmidt-probabilities}
\end{equation}
and write $\mathbf p=(p_0,p_1,p_2,p_3)$. Taking the moduli squared of
Eq.~\eqref{eq:kak-coefficients} gives
\begin{align}
    p_0 &=
    c_1^2c_2^2c_3^2
    +s_1^2s_2^2s_3^2,
    \nonumber\\
    p_1 &=
    s_1^2c_2^2c_3^2
    +c_1^2s_2^2s_3^2,
    \nonumber\\
    p_2 &=
    c_1^2s_2^2c_3^2
    +s_1^2c_2^2s_3^2,
    \nonumber\\
    p_3 &=
    c_1^2c_2^2s_3^2
    +s_1^2s_2^2c_3^2.
    \label{eq:schmidt-probabilities-cartan}
\end{align}

For arbitrary two-qubit unitaries, Schmitt, Piveteau and Sutter prove
that the optimal values of $\gamma$ under local operations and under
local operations assisted by classical communication coincide
\cite{Schmitt_2025}. Using $p_j=|u_j|^2$, their result can
be rewritten as
\begin{equation}
    \gamma
    =
    2\left(
        \sum_{j=0}^{3}\sqrt{p_j}
    \right)^2
    -1.
    \label{eq:gamma-schmidt}
\end{equation}
Equation~\eqref{eq:gamma-schmidt} therefore gives the optimal
single-gate \gls{QPD} extent. The quantity $\gamma$ should not be
confused with the literal number of circuit shots: for standard
quasiprobability estimation, the sampling overhead scales as
$\gamma^2$ for a single cut \cite{Schmitt_2025}.

The operator-Schmidt spectrum can also be characterised through its
R\'enyi entropies \cite{Musz_2013}. Using base-two logarithms, for
$\alpha>0$ and $\alpha\neq1$ we define
\begin{equation}
    H_\alpha(\mathbf p)
    =
    \frac{1}{1-\alpha}
    \log_2\!\left(
        \sum_j p_j^\alpha
    \right),
    \label{eq:renyi-definition}
\end{equation}
with $H_1$ understood by continuous extension as the Shannon entropy.
Setting $\alpha=1/2$ in Eq.~\eqref{eq:renyi-definition} and combining
the result with Eq.~\eqref{eq:gamma-schmidt} gives
\begin{equation}
    \frac{\gamma+1}{2}
    =
    2^{H_{1/2}(\mathbf p)}.
    \label{eq:gamma-renyi}
\end{equation}
Thus, the optimal two-qubit \gls{QPD} extent can be written exactly
as a R\'enyi-$1/2$ functional of its normalised operator-Schmidt
spectrum. This is a spectral reformulation of the known optimal
two-qubit cutting formula, rather than a new expression for the
optimum; related R\'enyi-$1/2$ extent formulae arise for bipartite pure
states in Ref.~\cite{https://doi.org/10.3929/ethz-b-000727956}. Its role here is to organise
the constrained extremal problem over physical two-qubit
operator-Schmidt spectra.

% ----------------------------------------------------------------
\subsection{Nonlocality quantities considered}
\label{subsec:nonlocality-quantities}
% ----------------------------------------------------------------

We compare $\gamma$ with five quantities that describe different
aspects of the nonlocal action of a two-qubit gate.

\subsubsection{Entangling power}

We use the linear-entropy entangling power introduced by Zanardi,
Zalka and Faoro, averaged uniformly over product inputs
\cite{Zanardi_2000},
\begin{equation}
    e_{\mathrm p}(U)
    =
    \int
    \mathrm d\psi_A\,\mathrm d\psi_B\,
    \left[
        1-\operatorname{Tr}\!\left(\rho_A^2\right)
    \right],
    \label{eq:ep-definition}
\end{equation}
where
\begin{equation}
    \rho_A
    =
    \operatorname{Tr}_B
    \left[
        U
        \left(
            |\psi_A\rangle\langle\psi_A|
            \otimes
            |\psi_B\rangle\langle\psi_B|
        \right)
        U^\dagger
    \right].
\end{equation}
In this normalisation, $0\leq e_{\mathrm p}\leq2/9$ for two-qubit
unitaries. For two qubits, Jonnadula \emph{et al.} use a
normalisation larger by a factor of three
\cite{Jonnadula_2020}. Their Cartan convention has parameters
$c_j=-2\theta_j$ relative to Eq.~\eqref{eq:cartan-unitary}.
Converting their two-qubit Cartan expression to the normalisation and
parameter convention used here, and simplifying, gives
\begin{equation}
    e_{\mathrm p}
    =
    \frac{1}{18}
    \left[
        3
        -\cos(4\theta_1)\cos(4\theta_2)
        -\cos(4\theta_2)\cos(4\theta_3)
        -\cos(4\theta_3)\cos(4\theta_1)
    \right].
    \label{eq:ep-cartan}
\end{equation}

\subsubsection{Gate typicality}

Gate typicality $g_{\mathrm t}$ was introduced by Jonnadula
\emph{et al.} as a complementary local-unitary invariant that
distinguishes, in particular, local gates from gates locally
equivalent to SWAP, both of which have zero entangling power
\cite{PhysRevA.95.040302}. We adopt the rescaled convention of their
later work \cite{Jonnadula_2020}, for which
$0\leq g_{\mathrm t}\leq1$, with the endpoints attained by local
gates and SWAP and its local equivalents, respectively. With the
same Cartan-parameter conversion used above, their two-qubit
expression becomes
\begin{equation}
    g_{\mathrm t}
    =
    \frac{1}{3}
    \sum_{j=1}^{3}
    \sin^2(2\theta_j).
    \label{eq:gt-cartan}
\end{equation}

\subsubsection{Operator entanglement}

Using the normalised operator-Schmidt spectrum introduced above, we
define the linear operator entanglement \cite{Jonnadula_2020}
\begin{equation}
    E_{\mathrm{op}}
    =
    1-\sum_{j=0}^{3}p_j^2,
    \qquad
    0\leq E_{\mathrm{op}}\leq\frac{3}{4}.
    \label{eq:eop-definition}
\end{equation}
For two qubits, solving the standard relations between operator
entanglement, entangling power and gate typicality in
Ref.~\cite{Jonnadula_2020}, and converting to the entangling-power
normalisation of Eq.~\eqref{eq:ep-definition}, gives
\begin{equation}
    E_{\mathrm{op}}
    =
    \frac{9}{8}e_{\mathrm p}
    +
    \frac{3}{4}g_{\mathrm t}.
    \label{eq:eop-ep-gt}
\end{equation}

\subsubsection{Schmidt strength}

Nielsen \emph{et al.} introduced the Schmidt strength as the Shannon
entropy of the normalised squared operator-Schmidt coefficients
\cite{Nielsen_2003}. In the normalised spectrum notation of
Eq.~\eqref{eq:schmidt-probabilities}, also used by Musz \emph{et al.}
\cite{Musz_2013}, and with base-two logarithms, this becomes
\begin{equation}
    K_{\mathrm{Sch}}
    =
    -\sum_{j=0}^{3}
    p_j\log_2 p_j,
    \qquad
    0\leq K_{\mathrm{Sch}}\leq2.
    \label{eq:ksch-definition}
\end{equation}
With this convention, $K_{\mathrm{Sch}}$ is measured in bits. We
also use the binary entropy
\begin{equation}
    h_2(x)
    =
    -x\log_2 x-(1-x)\log_2(1-x),
    \label{eq:binary-entropy}
\end{equation}
with the standard convention $0\log_2 0=0$.

\subsubsection{Maximum product-input concurrence}

The product-input entangling capacity of a two-qubit unitary can be
quantified by the maximum concurrence generated from a product state
\cite{Kraus_2001,Chefles_2005}. We denote this quantity by
\begin{equation}
    C_{\max}(U)
    =
    \max_{|\psi_A\rangle,|\psi_B\rangle}
    C\!\left(
        U|\psi_A\rangle|\psi_B\rangle
    \right),
    \qquad
    0\leq C_{\max}\leq1,
    \label{eq:cmax-definition}
\end{equation}
where $C$ is the two-qubit pure-state concurrence. Using the
invariance under $\theta_3\mapsto-\theta_3$ established above, the
product-entangling-capacity results of
Refs.~\cite{Kraus_2001,Chefles_2005} take the following form in our
Cartan convention:
\begin{equation}
    C_{\max}
    =
    \begin{cases}
        \sin\!\left[2(\theta_1+\theta_2)\right],
        &
        \theta_1+\theta_2<\pi/4,
        \\[1mm]
        1,
        &
        \theta_1+\theta_2\geq\pi/4,
        \quad
        \theta_2+|\theta_3|\leq\pi/4,
        \\[1mm]
        \sin\!\left[2(\theta_2+|\theta_3|)\right],
        &
        \theta_2+|\theta_3|>\pi/4.
    \end{cases}
    \label{eq:cmax-cartan}
\end{equation}
The middle branch is the perfect-entangler region: these gates can
transform at least one product input into a maximally entangled
two-qubit state \cite{PhysRevA.67.042313,Chefles_2005}.

Like $\gamma$, all five quantities are invariant under local pre- and
post-unitaries, but they capture distinct aspects of two-qubit
nonlocality. We therefore ask how sharply each constrains the lower and
upper extrema of $\gamma$.

% ================================================================
\section{Canonical Gate Families}
\label{sec:canonical-families}
% ================================================================

The six edges of the non-negative tetrahedron in
Figure~\ref{fig:cartan-tetrahedron} define the canonical
one-parameter families used below. Their Weyl-chamber geometry and
nonlocal properties have been characterised previously using
entangling power and local invariants \cite{Balakrishnan_2009}, and
later using operator-Schmidt structure and Schmidt strength
\cite{Balakrishnan_2011_QIP}; related comparisons of Schmidt strength
and linear operator entanglement were developed in
Ref.~\cite{Balakrishnan_2011_PRA}. Here we use these established edge
geometries as candidate and equality families in the \gls{QPD} extent
extremal problem. Table~\ref{tab:canonical-families}
gives their descriptive labels and roles in our Cartan convention
\cite{PhysRevA.67.042313}, with $0\leq\theta\leq\pi/4$. Five of these families attain at least one of the sharp envelopes
derived below. The remaining CNOT--iSWAP family is auxiliary to the
gate-typicality analysis and is spectrally degenerate with the
pair-pair family represented by CNOT--SWAP.

\begin{table}[t]
    \centering
    \small
    \caption{Canonical one-parameter families associated with the six
    edges of the non-negative tetrahedron. The final column
    summarises their role in the sharp-envelope results derived later
    in the paper.}
    \label{tab:canonical-families}
    \renewcommand{\arraystretch}{1.15}
    \begin{tabular}{@{}p{0.17\linewidth}p{0.20\linewidth}p{0.18\linewidth}p{0.32\linewidth}@{}}
        \toprule
        Family & Cartan path & Endpoint classes & Principal later role \\
        \midrule
        Controlled
        & $(\theta,0,0)$
        & $\mathbb I\rightarrow\mathrm{CNOT}$
        & Lower envelopes for $e_{\mathrm p}$ and $C_{\max}$; low-$g_{\mathrm t}$,
        low-$E_{\mathrm{op}}$ and low-$K_{\mathrm{Sch}}$ lower branches. \\

        XY/iSWAP
        & $(\theta,\theta,0)$
        & $\mathbb I\rightarrow\mathrm{iSWAP}$
        & Middle-$g_{\mathrm t}$ upper branch. \\

        Fractional SWAP
        & $(\theta,\theta,\theta)$
        & $\mathbb I\rightarrow\mathrm{SWAP}$
        & Upper envelopes for $E_{\mathrm{op}}$ and $K_{\mathrm{Sch}}$;
        low-$g_{\mathrm t}$ upper and high-$g_{\mathrm t}$ lower branches. \\

        CNOT--iSWAP
        & $(\pi/4,\theta,0)$
        & $\mathrm{CNOT}\rightarrow\mathrm{iSWAP}$
        & Auxiliary boundary family in the $g_{\mathrm t}$ extremality
        analysis; spectrally degenerate with the pair-pair family. \\

        iSWAP--SWAP
        & $(\pi/4,\pi/4,\theta)$
        & $\mathrm{iSWAP}\rightarrow\mathrm{SWAP}$
        & $\gamma=7$ upper envelopes for $e_{\mathrm p}$ and $C_{\max}$;
        high-$g_{\mathrm t}$ upper branch. \\

        CNOT--SWAP
        & $(\pi/4,\theta,\theta)$
        & $\mathrm{CNOT}\rightarrow\mathrm{SWAP}$
        & Middle-$g_{\mathrm t}$ lower branch; high-$E_{\mathrm{op}}$ and
        high-$K_{\mathrm{Sch}}$ lower branches. \\
        \bottomrule
    \end{tabular}
\end{table}

\begin{figure}[t]
    \centering
    \includegraphics[width=0.8\linewidth]{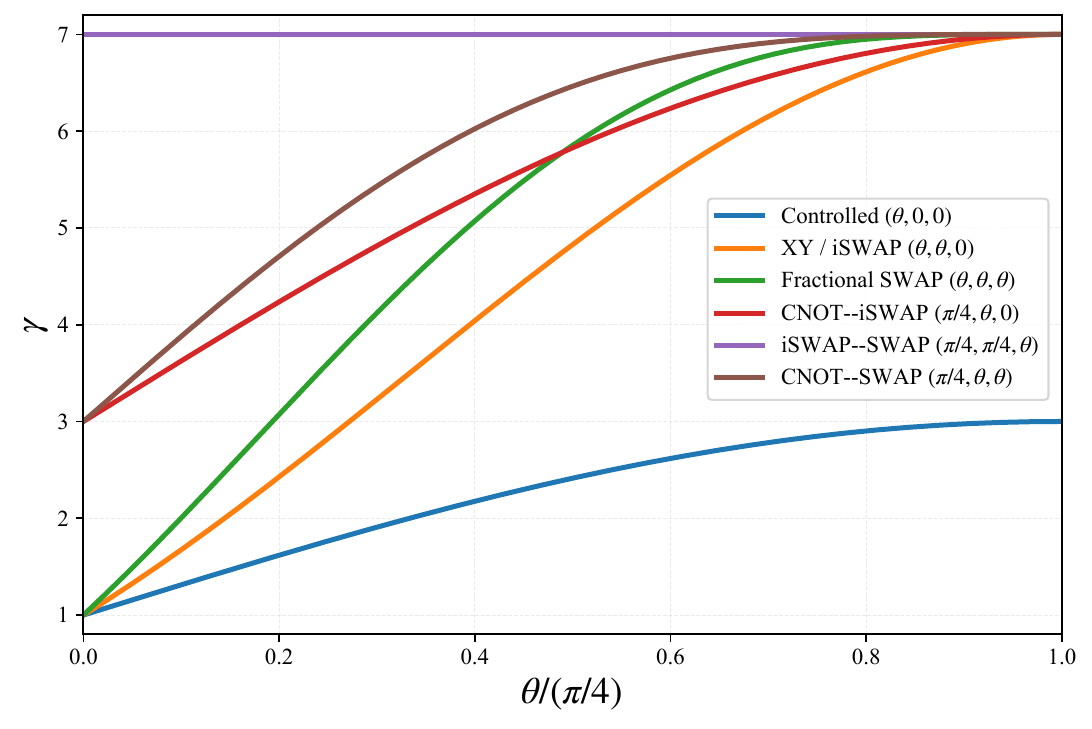}
    \caption{Optimal \gls{QPD} extent along the six canonical edge families.
    The curves are obtained by substituting the Cartan paths in
    Table~\ref{tab:canonical-families} into
    Eq.~\eqref{eq:gamma-schmidt}. Each family uses
    $0\leq\theta\leq\pi/4$, and the horizontal axis shows
    $\theta/(\pi/4)$.}
    \label{fig:canonical-edge-gamma}
\end{figure}

The trajectories already show the separation between cutting cost and
other notions of nonlocal action. The controlled family rises from
$\gamma=1$ at the identity to $\gamma=3$ at CNOT, whereas the
iSWAP--SWAP edge has the uniform spectrum
$\mathbf p=(1/4,1/4,1/4,1/4)$ and hence $\gamma=7$ throughout despite
variation in $e_{\mathrm p}$, $g_{\mathrm t}$ and $C_{\max}$.
Appendix~\ref{app:canonical-families} gives the spectra and closed-form
expressions for all six families.

% ================================================================
\section{Cutting Cost versus Gate-Action Measures}
\label{sec:gate-action}
% ================================================================

Entangling power and gate typicality describe complementary aspects of
two-qubit gate action. Each defines exact lower and upper extrema for
the \gls{QPD} extent rather than determining it uniquely. Throughout,
the shaded band between these curves is an \emph{outer-envelope band}:
every physical Cartan point projects into it or onto its boundary, and
both boundaries are attained at every descriptor value, but the band
need not be pointwise filled. Indeed,
Eqs.~\eqref{eq:ep-cartan} and \eqref{eq:cmax-cartan} imply that at
$e_{\mathrm p}=0$ and $C_{\max}=0$ the only local-equivalence classes
are the identity and SWAP classes, so the attainable \gls{QPD} extents
at either slice are $\{1,7\}$ rather than the full interval
$[1,7]$.

\begin{theorem}[Gate-action envelopes]
\label{thm:gate-action-envelopes}
For any two-qubit unitary, the sharp extrema are as follows.

\emph{(i) Entangling power.}
For $0\leq e_{\mathrm p}\leq2/9$,
\begin{equation}
    1+3\sqrt{2e_{\mathrm p}}
    \leq
    \gamma
    \leq
    7.
    \label{eq:ep-envelope}
\end{equation}
The lower boundary is attained by the controlled family
$(\theta,0,0)$, and the upper boundary by the iSWAP--SWAP family
$(\pi/4,\pi/4,\theta)$.

\emph{(ii) Gate typicality.}
For $0\leq g_{\mathrm t}\leq1$, the lower envelope is
\begin{equation}
    \gamma_{\min}(g_{\mathrm t})
    =
    \begin{cases}
        1+2\sqrt{3g_{\mathrm t}},
        &
        0\leq g_{\mathrm t}\leq\frac13,
        \\[1mm]
        3+2\sqrt{(3g_{\mathrm t}-1)(5-3g_{\mathrm t})},
        &
        \frac13\leq g_{\mathrm t}\leq\frac23,
        \\[1mm]
        1+3g_{\mathrm t}
        +3\sqrt{g_{\mathrm t}(4-3g_{\mathrm t})},
        &
        \frac23\leq g_{\mathrm t}\leq1,
    \end{cases}
    \label{eq:gt-lower-envelope}
\end{equation}
and the upper envelope is
\begin{equation}
    \gamma_{\max}(g_{\mathrm t})
    =
    \begin{cases}
        1+3g_{\mathrm t}
        +3\sqrt{g_{\mathrm t}(4-3g_{\mathrm t})},
        &
        0\leq g_{\mathrm t}\leq\frac49,
        \\[1mm]
        1+3g_{\mathrm t}
        +4\sqrt{\frac{3g_{\mathrm t}}{2}},
        &
        \frac49\leq g_{\mathrm t}\leq\frac23,
        \\[1mm]
        7,
        &
        \frac23\leq g_{\mathrm t}\leq1.
    \end{cases}
    \label{eq:gt-upper-envelope}
\end{equation}
As $g_{\mathrm t}$ increases, the lower boundary is attained
successively by the controlled, CNOT--SWAP and fractional SWAP
families, while the upper boundary is attained by the fractional SWAP,
XY/iSWAP and iSWAP--SWAP families.
\end{theorem}

The proof of Theorem~\ref{thm:gate-action-envelopes}, including the
global boundary classification and equality conditions, is deferred to
Appendix~\ref{app:gate-action-extremality}.

\subsection{Entangling power}
\label{subsec:gamma-ep}

The entangling-power envelope is strongly asymmetric: the controlled
family raises the lower boundary from $\gamma=1$ to $\gamma=3$, while
the iSWAP--SWAP family keeps the upper boundary at $\gamma=7$ across
the full interval $0\leq e_{\mathrm p}\leq2/9$. Accordingly,
$e_{\mathrm p}=0$ permits the identity and SWAP values
$\gamma=1$ and $7$, while at $e_{\mathrm p}=2/9$ CNOT and iSWAP give
$\gamma=3$ and $7$. Entangling power therefore provides a nontrivial
lower constraint but does not improve the universal upper bound.

\begin{figure}[t]
    \centering
    \includegraphics[width=0.8\linewidth]{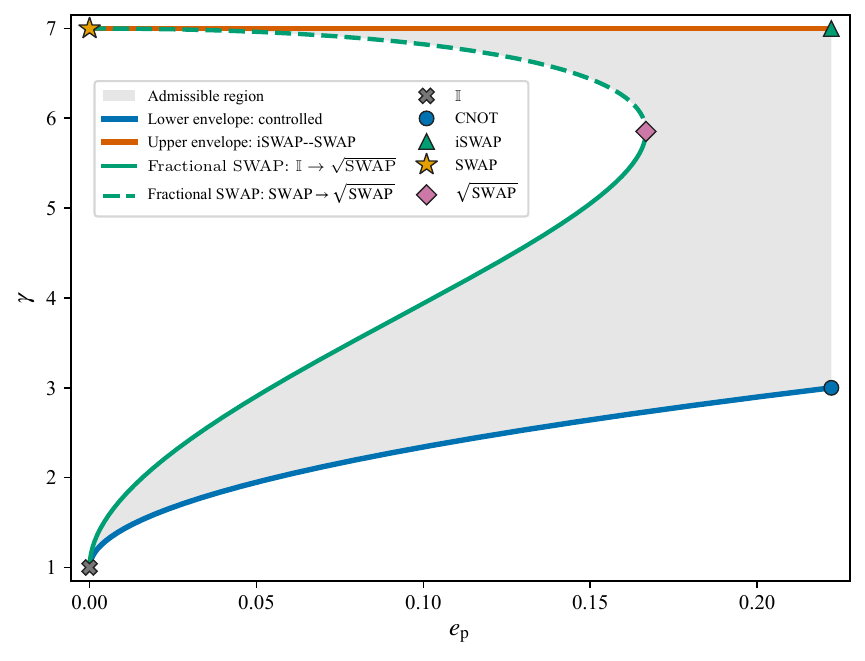}
    \caption{Optimal \gls{QPD} extent versus entangling power. The solid
    curves are the exact envelopes in Eq.~\eqref{eq:ep-envelope},
    attained by the controlled and iSWAP--SWAP families. The shaded
    area is the sharp outer-envelope band: every physical Cartan point
    projects into it or onto its boundary, but the interior need not
    be pointwise realised.}
    \label{fig:gamma-entangling-power}
\end{figure}

\subsection{Gate typicality}
\label{subsec:gamma-gt}

Gate typicality gives piecewise lower and upper boundaries. The lower
equality family changes at $g_{\mathrm t}=1/3$ and $2/3$; the upper
boundary has the additional nontrivial fractional-SWAP/XY crossing at
$g_{\mathrm t}=4/9$, where
\begin{equation}
    \gamma
    =
    \frac{7}{3}
    +
    \frac{4\sqrt{6}}{3},
    \label{eq:gt-four-ninths-gamma}
\end{equation}
producing a derivative kink. At $g_{\mathrm t}=2/3$, the XY/iSWAP
branch reaches $\gamma=7$ and joins the iSWAP--SWAP plateau. Unlike
entangling power, the endpoints fix the extent:
$g_{\mathrm t}=0$ gives $\gamma=1$ and $g_{\mathrm t}=1$ gives
$\gamma=7$, although the lower and upper envelopes remain separated
in the interior.

\begin{figure}[t]
    \centering
    \includegraphics[width=0.8\linewidth]{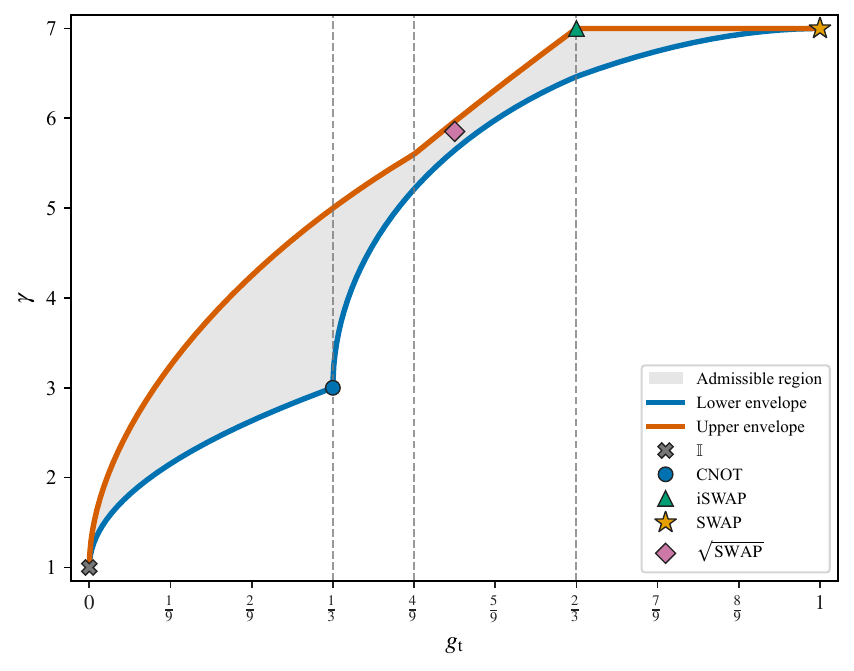}
    \caption{Optimal \gls{QPD} extent versus gate typicality. The solid
    curves are the exact lower and upper envelopes in
    Eqs.~\eqref{eq:gt-lower-envelope} and
    \eqref{eq:gt-upper-envelope}. The upper boundary switches from the
    fractional SWAP to the XY/iSWAP family at $g_{\mathrm t}=4/9$ and
    reaches the uniform-spectrum $\gamma=7$ plateau at
    $g_{\mathrm t}=2/3$. The shaded area is the corresponding sharp
    outer-envelope band.}
    \label{fig:gamma-gate-typicality}
\end{figure}

% ================================================================
\section{Cutting Cost versus Operator-Schmidt Entanglement}
\label{sec:spectral-entanglement}
% ================================================================

Operator entanglement and Schmidt strength are defined directly from
the normalised operator-Schmidt spectrum. Physical two-qubit-unitary
spectra occupy only a subset of the four-component probability simplex,
which sharpens both lower boundaries.

\subsection{Physical two-qubit operator-Schmidt spectra}
\label{subsec:physical-schmidt-spectra}

To distinguish an ordered spectrum from the fixed Pauli-labelled
probabilities $p_0,\ldots,p_3$ of Section~\ref{subsec:schmidt-gamma},
write
\begin{equation}
    p_{(1)}
    \geq
    p_{(2)}
    \geq
    p_{(3)}
    \geq
    p_{(4)}
    \geq0,
    \qquad
    \sum_{j=1}^{4}p_{(j)}=1.
    \label{eq:ordered-physical-spectrum}
\end{equation}
The absence of operator-Schmidt rank three for two-qubit
unitaries is known from earlier work
\cite{Dur_2002,Tyson_2003}. For the present extremal problem, the
stronger useful input is the ordered physicality condition obtained
from the unistochastic characterisation of two-qubit
operator-Schmidt spectra in Ref.~\cite{Musz_2013}. After ordering, the
relevant restriction takes the following simple form.

\begin{lemma}[Physical two-qubit operator-Schmidt constraint]
\label{lem:physical-schmidt-spectrum}
Every two-qubit-unitary operator-Schmidt spectrum satisfies
\begin{equation}
    p_{(1)}p_{(4)}
    \geq
    p_{(2)}p_{(3)}.
    \label{eq:physical-schmidt-constraint}
\end{equation}
Consequently, operator-Schmidt rank three is impossible.
\end{lemma}

Appendix~\ref{app:physical-schmidt-spectra} gives the derivation,
rank consequences and boundary factorisation. This restriction is why
generic probability-simplex entropy bounds need not be sharp.

\begin{theorem}[Operator-Schmidt spectral envelopes]
\label{thm:spectral-envelopes}
For any two-qubit unitary, the following lower and upper envelopes are
sharp.

\emph{(i) Operator entanglement.}
For $0\leq E_{\mathrm{op}}\leq3/4$,
\begin{equation}
    \gamma_{\min}(E_{\mathrm{op}})
    =
    \begin{cases}
        1+2\sqrt{2E_{\mathrm{op}}},
        &
        0\leq E_{\mathrm{op}}\leq\frac12,
        \\[1mm]
        3+4\sqrt{4E_{\mathrm{op}}-2},
        &
        \frac12\leq E_{\mathrm{op}}\leq\frac34.
    \end{cases}
    \label{eq:eop-lower-envelope}
\end{equation}
The lower equality spectrum changes at $E_{\mathrm{op}}=1/2$ from
rank-two controlled spectra to pair-pair spectra, for which
CNOT--SWAP is a canonical representative. The upper boundary is
generated by the fractional SWAP family and can be written as
\begin{equation}
    \gamma_{\max}(E_{\mathrm{op}})
    =
    4-3q_E
    +
    3\sqrt{(1-q_E)(1+3q_E)},
    \qquad
    q_E
    =
    \sqrt{1-\frac{4}{3}E_{\mathrm{op}}}.
    \label{eq:eop-upper-envelope}
\end{equation}

\emph{(ii) Schmidt strength.}
Let $z(u)$ denote the unique $z\in[1/2,1]$ satisfying
$h_2(z)=u$, where $h_2$ is defined in
Eq.~\eqref{eq:binary-entropy}. Then
\begin{equation}
    \gamma_{\min}(K_{\mathrm{Sch}})
    =
    \begin{cases}
        1
        +
        4\sqrt{
            z(K_{\mathrm{Sch}})
            \left[1-z(K_{\mathrm{Sch}})\right]
        },
        &
        0\leq K_{\mathrm{Sch}}\leq1,
        \\[2mm]
        3
        +
        8\sqrt{
            z(K_{\mathrm{Sch}}-1)
            \left[1-z(K_{\mathrm{Sch}}-1)\right]
        },
        &
        1\leq K_{\mathrm{Sch}}\leq2.
    \end{cases}
    \label{eq:ksch-lower-envelope}
\end{equation}
The lower equality spectrum changes at
$K_{\mathrm{Sch}}=1$ bit from rank-two controlled spectra to
pair-pair spectra, for which CNOT--SWAP is a canonical
representative.

\begin{samepage}
The upper boundary is generated by the fractional SWAP family and is
most compactly stated parametrically as
\begin{align}
    K_{\mathrm{FS}}(x)
    &=
    h_2\!\left(\frac{3x}{4}\right)
    +
    \frac{3x}{4}\log_2 3,
    \nonumber\\
    \gamma_{\mathrm{FS}}(x)
    &=
    1+3x+3\sqrt{x(4-3x)},
    \qquad
    0\leq x\leq1.
    \label{eq:ksch-upper-envelope-parametric}
\end{align}
The function $K_{\mathrm{FS}}(x)$ is strictly increasing on this
interval, so the parametric curve defines a unique
$\gamma_{\max}(K_{\mathrm{Sch}})$.
\end{samepage}
\end{theorem}

Appendix~\ref{app:entropy-envelope-derivations} gives the entropy
extremality provenance and the derivation of the sharp physical
envelopes in Theorem~\ref{thm:spectral-envelopes}.

\subsection{Operator entanglement}
\label{subsec:gamma-eop}

Below $E_{\mathrm{op}}=1/2$, the lower boundary is the rank-two
controlled family; above it, the minimum follows the pair-pair locus
represented by CNOT--SWAP and reaches the uniform spectrum at
$E_{\mathrm{op}}=3/4$. Fractional SWAP generates the upper boundary
throughout and physically realises the corresponding generic
four-outcome $1+3$ entropy extremum.

For comparison, R\'enyi-order monotonicity gives
\begin{equation}
    \gamma
    \geq
    \frac{1+E_{\mathrm{op}}}{1-E_{\mathrm{op}}}.
    \label{eq:eop-generic-renyi-bound}
\end{equation}
This generic probability-space bound is valid but touches the sharp
physical lower boundary only at $E_{\mathrm{op}}=0$, $1/2$ and
$3/4$. The uniform rank-three equality spectrum at
$E_{\mathrm{op}}=2/3$ is excluded by
Lemma~\ref{lem:physical-schmidt-spectrum}, so the physical spectrum
constraint is essential for sharpness.

\begin{figure}[t]
    \centering
    \includegraphics[width=0.8\linewidth]{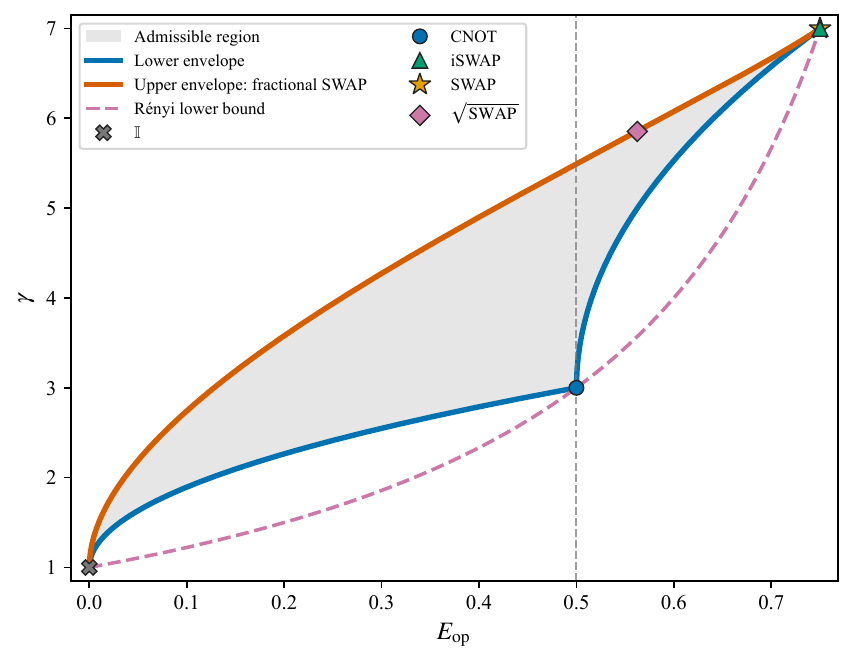}
    \caption{Optimal \gls{QPD} extent versus operator entanglement. The
    solid curves are the exact physical two-qubit-unitary envelopes in
    Eqs.~\eqref{eq:eop-lower-envelope} and
    \eqref{eq:eop-upper-envelope}. The lower equality spectrum changes
    from controlled rank-two to pair-pair at $E_{\mathrm{op}}=1/2$,
    while fractional SWAP generates the upper boundary. The shaded
    area is the corresponding sharp outer-envelope band.}
    \label{fig:gamma-operator-entanglement}
\end{figure}

\subsection{Schmidt strength}
\label{subsec:gamma-ksch}

Schmidt strength has the same lower-family switch in entropy form.
Controlled rank-two spectra minimise the \gls{QPD} extent up to
$K_{\mathrm{Sch}}=1$, where CNOT has $\gamma=3$; the pair-pair locus
then reaches the uniform spectrum at $K_{\mathrm{Sch}}=2$ and
$\gamma=7$. Fractional SWAP again gives the complete upper boundary.
Because $K_{\mathrm{FS}}(x)$ is strictly increasing but has no useful
elementary inverse, Eq.~\eqref{eq:ksch-upper-envelope-parametric} is
most naturally left parametric.

The generic R\'enyi-order inequality
\begin{equation}
    \gamma
    \geq
    2^{K_{\mathrm{Sch}}+1}-1
    \label{eq:ksch-generic-renyi-bound}
\end{equation}
touches the sharp physical lower envelope only at
$(K_{\mathrm{Sch}},\gamma)=(0,1)$, $(1,3)$ and $(2,7)$, the
uniform-support spectra of ranks one, two and four. At intermediate
Schmidt strength the physical spectrum geometry raises the sharp lower
bound above this generic inequality.

\begin{figure}[t]
    \centering
    \includegraphics[width=0.8\linewidth]{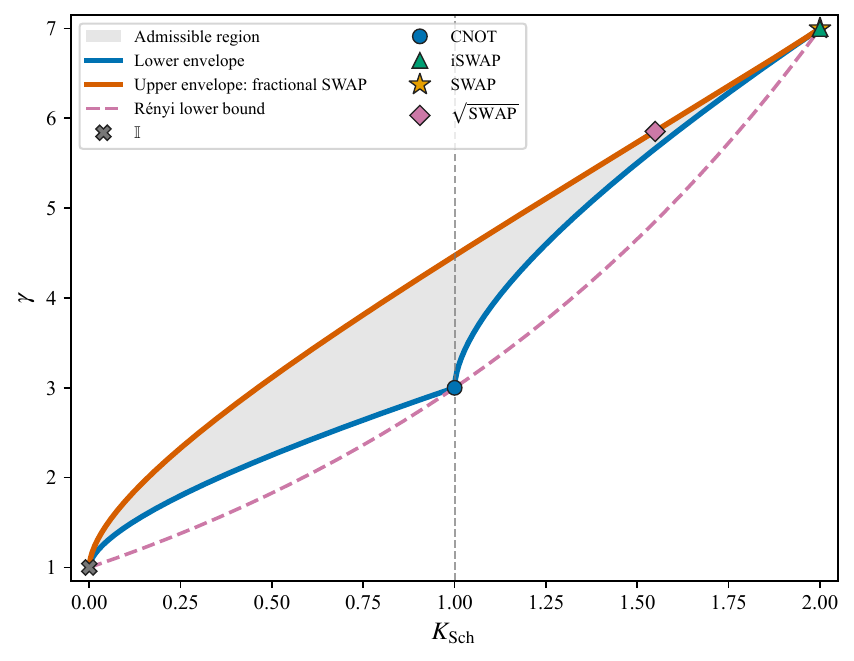}
    \caption{Optimal \gls{QPD} extent versus Schmidt strength. The solid
    curves are the exact physical two-qubit-unitary envelopes of
    Theorem~\ref{thm:spectral-envelopes}. The lower equality spectrum
    changes from controlled rank-two to pair-pair at one bit, while
    fractional SWAP generates the complete upper boundary. The shaded
    area is the corresponding sharp outer-envelope band.}
    \label{fig:gamma-schmidt-strength}
\end{figure}

% ================================================================
\section{Cutting Cost versus Entangling Capacity}
\label{sec:entangling-capacity}
% ================================================================

The maximum product-input concurrence $C_{\max}$ of
Eq.~\eqref{eq:cmax-definition} gives the simplest of the five sharp
envelopes.

\begin{theorem}[Maximum-concurrence envelope]
\label{thm:cmax-envelope}
For every two-qubit unitary,
\begin{equation}
    1+2C_{\max}
    \leq
    \gamma
    \leq
    7,
    \qquad
    0\leq C_{\max}\leq1.
    \label{eq:cmax-envelope}
\end{equation}
Both bounds are realised for every fixed $C_{\max}$. The controlled
family attains the lower boundary, while the iSWAP--SWAP family
attains the upper boundary.
\end{theorem}

The lower inequality in Theorem~\ref{thm:cmax-envelope} is also a
two-qubit specialisation of the general state-conversion lower bound
for channel quasiprobability extent in
Ref.~\cite{https://doi.org/10.3929/ethz-b-000727956}. The additional
content here is the pair of sharp fixed-$C_{\max}$ envelopes: both
boundaries are shown to be realised for every $C_{\max}$, with
explicit equality families. A direct proof and the relevant Cartan
geometry are given in
Appendix~\ref{app:cmax-perfect-entanglers}.

\subsection{Maximum product-input concurrence}
\label{subsec:gamma-cmax}

Along the controlled family, $C_{\max}$ rises from zero to one while
$\gamma$ increases linearly from $1$ to $3$; along iSWAP--SWAP,
$\gamma=7$ while $C_{\max}$ spans the same interval in reverse. Thus, at $C_{\max}=0$, the only attainable values are the identity
and SWAP values $\gamma=1$ and $7$, whereas $C_{\max}=1$ contains
CNOT and iSWAP at $\gamma=3$ and $7$ and, as shown below, the full
interval between them. Maximum concurrence therefore gives an exact
linear lower constraint but no improvement over the universal upper
bound.

\begin{figure}[t]
    \centering
    \includegraphics[width=0.8\linewidth]{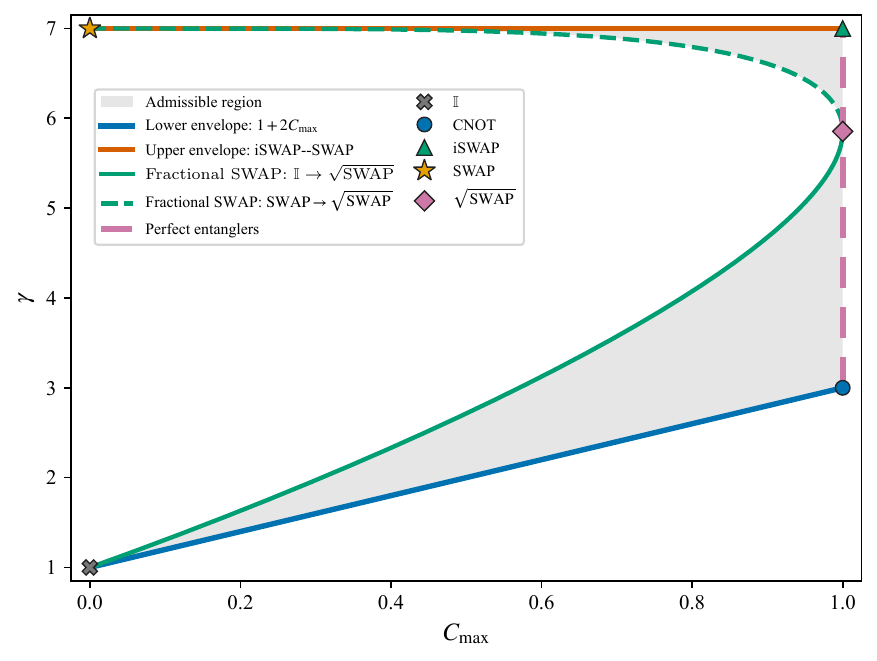}
    \caption{Optimal \gls{QPD} extent versus maximum product-input
    concurrence. The solid boundaries are the exact envelopes of
    Theorem~\ref{thm:cmax-envelope}, attained by the controlled and
    iSWAP--SWAP families. The shaded area is the corresponding sharp
    outer-envelope band; intervening values within the band are not
    asserted to be attainable.}
    \label{fig:gamma-cmax}
\end{figure}

\subsection{Perfect entanglers}
\label{subsec:perfect-entanglers}

By the concurrence criterion in Eq.~\eqref{eq:cmax-cartan}, a
two-qubit gate is a perfect entangler exactly when $C_{\max}=1$.
Theorem~\ref{thm:cmax-envelope} therefore gives the following sharp
consequence.

\begin{corollary}[Minimum \gls{QPD} extent of a perfect entangler]
\label{cor:perfect-entangler-cost}
Every two-qubit perfect entangler satisfies $\gamma\geq3$, and the
bound is attained by CNOT. Hence
\begin{equation}
    \gamma_{\min}^{\mathrm{PE}}
    =
    3.
    \label{eq:perfect-entangler-min}
\end{equation}
\end{corollary}

At the opposite extreme, iSWAP is also a perfect entangler and has
$\gamma=7$. Thus the perfect-entangler class spans the sharp range
$3\leq\gamma\leq7$: perfect entangling capability alone does not
determine the optimal cutting cost. The corresponding Cartan geometry
is described in Appendix~\ref{app:cmax-perfect-entanglers}.

% ================================================================
\section{Discussion}
\label{sec:discussion}
% ================================================================

To compare the five pairs of sharp envelopes on a common horizontal scale,
define
\begin{align}
    \widetilde e_{\mathrm p}
    &=
    \frac{9}{2}e_{\mathrm p},
    &
    \widetilde g_{\mathrm t}
    &=
    g_{\mathrm t},
    &
    \widetilde E_{\mathrm{op}}
    &=
    \frac{4}{3}E_{\mathrm{op}},
    \nonumber\\
    \widetilde K_{\mathrm{Sch}}
    &=
    \frac12K_{\mathrm{Sch}},
    &
    \widetilde C_{\max}
    &=
    C_{\max},
    \label{eq:normalised-descriptors}
\end{align}
so that every normalised descriptor lies in $[0,1]$, while
$1\leq\gamma\leq7$ is unchanged. This rescaling compares envelope
shapes and separations only; equal normalised values of different
descriptors are not physically equivalent.

At the zero endpoint, $g_{\mathrm t}$, $E_{\mathrm{op}}$ and
$K_{\mathrm{Sch}}$ force $\gamma=1$, whereas for
$e_{\mathrm p}=0$ and $C_{\max}=0$ the sharp lower and upper extrema
are $\gamma=1$ and $\gamma=7$. In the latter two cases the fixed-value
attainable sets are only $\{1,7\}$, realised by the identity and SWAP,
so the vertical interval between the envelopes is not filled. At the
opposite endpoint, maximal $g_{\mathrm t}$, $E_{\mathrm{op}}$ and
$K_{\mathrm{Sch}}$ force the uniform spectrum and $\gamma=7$, while
maximal $e_{\mathrm p}$ and $C_{\max}$ have sharp extrema
$3\leq\gamma\leq7$; along CNOT--iSWAP every value in this latter
interval is realised. Thus the first three descriptors fix the extent
at both extrema, while the two direct entangling-capability measures
retain distinct sharp extrema.

For interior values, none of the five descriptors determines
$\gamma$ uniquely. The envelope separation measures the uncertainty
left by the extremal constraints; it is not, in general, the length of
the physically attainable set of $\gamma$ values. For the two
constant-upper-bound cases, the normalised coordinate $q\in[0,1]$
gives
\begin{align}
    1+2\sqrt q
    &\leq
    \gamma
    \leq
    7
    &&
    \text{for entangling power},
    \nonumber\\
    1+2q
    &\leq
    \gamma
    \leq
    7
    &&
    \text{for maximum concurrence}.
    \label{eq:normalised-ep-cmax-comparison}
\end{align}
Entangling power therefore gives the stronger lower constraint for
$0<q<1$, although both retain the maximal upper bound. The
$g_{\mathrm t}$, $E_{\mathrm{op}}$ and $K_{\mathrm{Sch}}$
outer-envelope bands instead collapse at both endpoints and open only
in the interior. Their envelope separations cross as the normalised
coordinate varies, so there is no single ordering of the five
descriptors by extremal constraint strength.

\begin{figure}[t]
    \centering
    \begin{subfigure}[t]{0.497\linewidth}
        \centering
        \includegraphics[width=\linewidth]{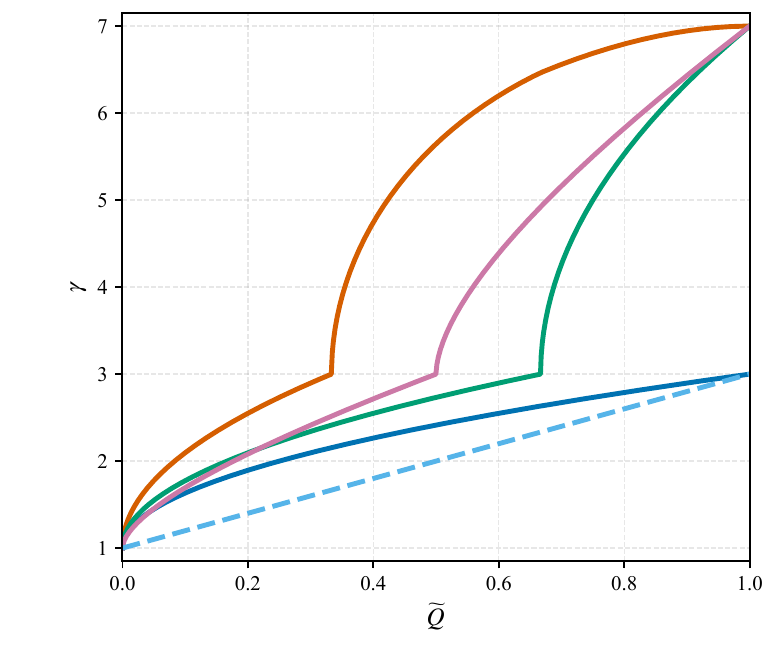}
        \caption{Lower envelopes.}
        \label{fig:normalised-envelope-comparison-lower}
    \end{subfigure}\hspace{0.006\linewidth}%
    \begin{subfigure}[t]{0.497\linewidth}
        \centering
        \includegraphics[width=\linewidth]{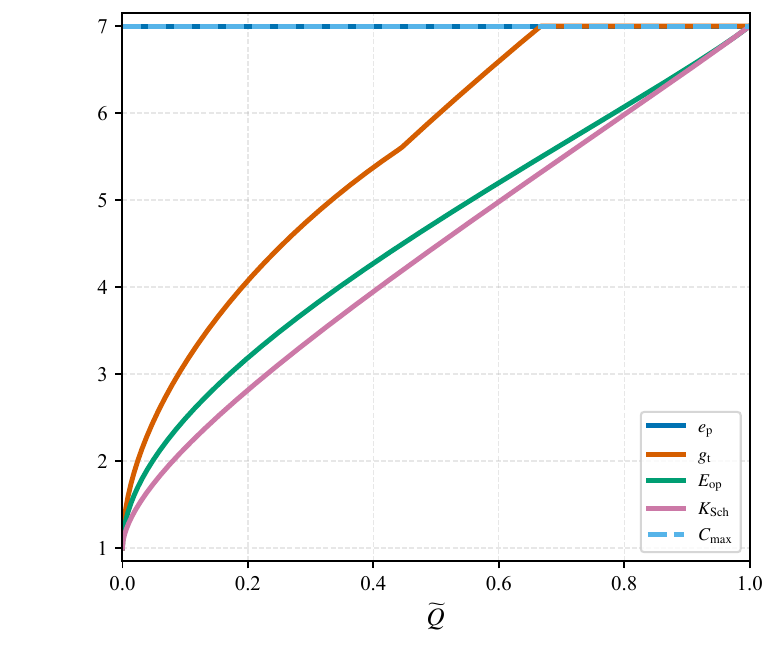}
        \caption{Upper envelopes.}
        \label{fig:normalised-envelope-comparison-upper}
    \end{subfigure}
    \caption{Comparison of the five pairs of exact sharp envelopes
    after normalising each descriptor to the unit interval according
    to Eq.~\eqref{eq:normalised-descriptors}. Panels (a) and (b) show
    the exact lower and upper envelopes, respectively, while the
    vertical axis retains the physical range $1\leq\gamma\leq7$.
    Read together, the vertical separation between corresponding
    curves is the outer-envelope separation at that normalised
    descriptor value, without implying that every intervening
    \gls{QPD} extent is physically attainable. The normalisation is for
    visual comparison only and does not identify equal normalised
    values of different descriptors.}
    \label{fig:normalised-envelope-comparison}
\end{figure}

All sharp boundaries are generated by a small set of nonlocal
geometries. The controlled family gives the low-cost branches at fixed
$e_{\mathrm p}$, $C_{\max}$ and low spectral entropy; the lower
spectral envelopes then move to the pair-pair locus represented by
CNOT--SWAP. Fractional SWAP supplies both spectral upper envelopes and
two gate-typicality branches, XY/iSWAP supplies the remaining
nontrivial $g_{\mathrm t}$ upper branch, and the uniform-spectrum
iSWAP--SWAP edge generates every $\gamma=7$ plateau. The recurring
structures are therefore rank-two, pair-pair, the symmetric
binary-product XY/iSWAP spectrum, $1+3$, and uniform spectra.

Equation~\eqref{eq:gamma-renyi} explains why no single familiar
descriptor is a universal proxy for cutting cost: $\gamma$ depends on
the R\'enyi-$1/2$ operator-Schmidt functional, while the five
descriptors constrain different scalar features of the same gate.
Even the spectral quantities $E_{\mathrm{op}}$ and
$K_{\mathrm{Sch}}$ do not determine $H_{1/2}$. Likewise,
$(e_{\mathrm p},g_{\mathrm t})$ fixes $E_{\mathrm{op}}$ through
Eq.~\eqref{eq:eop-ep-gt}, although neither component does so alone.
The one-dimensional envelopes quantify the resulting freedom.

The present results concern exact cutting of a single ideal two-qubit
unitary. A complementary extension is to characterise the topology and
connected components of the attainable sets inside the present
one-descriptor outer-envelope bands. Other natural extensions include
joint descriptor constraints, higher-dimensional bipartite gates,
multiple cut gates, and approximate or noisy decompositions. These
respectively require higher-dimensional feasible-set analyses,
extensions of the physical-spectrum analysis, compositional and
cut-placement considerations, and cost--bias or implementation
trade-offs beyond the present setting.

% ================================================================

\section{Conclusion}
\label{sec:conclusion}
% ================================================================

None of the five descriptors considered here determines the optimal
two-qubit \gls{QPD} extent generically, but for each we determine the
complete sharp lower and upper envelopes. Starting from the known
optimal cutting formula, we recast $\gamma$ as a R\'enyi-$1/2$
functional of the normalised operator-Schmidt spectrum and used the
physical two-qubit spectrum constraints to sharpen generic entropy
bounds.

The resulting envelopes include the complete piecewise
gate-typicality envelopes and physically sharpened
operator-entanglement and Schmidt-strength bounds. For maximum
concurrence, the general state-conversion argument of
Ref.~\cite{https://doi.org/10.3929/ethz-b-000727956} gives the lower
inequality, while the equality analysis here establishes both sharp
fixed-$C_{\max}$ boundaries in
$1+2C_{\max}\leq\gamma\leq7$, with each boundary realised for every
$C_{\max}$. The perfect-entangler range $3\leq\gamma\leq7$ follows
directly.

Across all five descriptor comparisons, the extremal boundaries
repeatedly select the same small set of Cartan families. This recurrence reflects the
underlying physical operator-Schmidt geometry: familiar scalar
descriptors probe different projections of that structure, whereas the
R\'enyi-$1/2$ functional determines the cutting cost itself.

% ================================================================

\clearpage
\section*{Appendices}
\appendix
% ================================================================

\section{Cartan Formulas and Canonical Families}
\label{app:canonical-families}

This appendix collects the closed-form expressions for the six
one-parameter families introduced in
Table~\ref{tab:canonical-families}. Unless stated otherwise, the
results follow by direct substitution of the corresponding Cartan
path into Eqs.~\eqref{eq:schmidt-probabilities-cartan},
\eqref{eq:gamma-schmidt}, \eqref{eq:ep-cartan},
\eqref{eq:gt-cartan}, \eqref{eq:eop-definition},
\eqref{eq:ksch-definition}, and \eqref{eq:cmax-cartan}. Since the
quantities considered here are determined by the Cartan parameters
and the normalised operator-Schmidt spectrum, we report
$\mathbf p$ directly rather than repeating the intermediate complex
coefficients $u_j$. We use
\begin{equation}
    c=\cos\theta,
    \qquad
    s=\sin\theta,
    \qquad
    r=\sin(2\theta),
    \qquad
    0\leq\theta\leq\frac{\pi}{4},
    \label{eq:appA-common-parameters}
\end{equation}
and use the binary entropy $h_2$ defined in
Eq.~\eqref{eq:binary-entropy}.

\subsection{Controlled family}
\label{app:family-controlled}

For the controlled family $(\theta,0,0)$, the normalised
operator-Schmidt spectrum is
\begin{equation}
    \mathbf p_{\mathrm{ctrl}}
    =
    \left(
        c^2,\,
        s^2,\,
        0,\,
        0
    \right).
    \label{eq:appA-controlled-spectrum}
\end{equation}
The corresponding quantities are
\begin{align}
    \gamma_{\mathrm{ctrl}}
    &=
    1+2r,
    \nonumber\\
    e_{\mathrm p}^{\mathrm{ctrl}}
    &=
    \frac{2}{9}r^2,
    \nonumber\\
    g_{\mathrm t}^{\mathrm{ctrl}}
    &=
    \frac{1}{3}r^2,
    \nonumber\\
    E_{\mathrm{op}}^{\mathrm{ctrl}}
    &=
    \frac{1}{2}r^2,
    \nonumber\\
    K_{\mathrm{Sch}}^{\mathrm{ctrl}}
    &=
    h_2(s^2),
    \nonumber\\
    C_{\max}^{\mathrm{ctrl}}
    &=
    r.
    \label{eq:appA-controlled-quantities}
\end{align}
The interior of this edge has operator-Schmidt rank two; the identity
endpoint has rank one.

\subsection{XY/iSWAP family}
\label{app:family-xy}

For the XY/iSWAP family $(\theta,\theta,0)$,
\begin{equation}
    \mathbf p_{\mathrm{XY}}
    =
    \left(
        c^4,\,
        c^2s^2,\,
        c^2s^2,\,
        s^4
    \right)
    =
    (c^2,s^2)\otimes(c^2,s^2).
    \label{eq:appA-xy-spectrum}
\end{equation}
The \gls{QPD} extent and the remaining non-piecewise quantities are
\begin{align}
    \gamma_{\mathrm{XY}}
    &=
    2(1+r)^2-1,
    \nonumber\\
    e_{\mathrm p}^{\mathrm{XY}}
    &=
    \frac{2}{9}r^2(2-r^2),
    \nonumber\\
    g_{\mathrm t}^{\mathrm{XY}}
    &=
    \frac{2}{3}r^2,
    \nonumber\\
    E_{\mathrm{op}}^{\mathrm{XY}}
    &=
    r^2-\frac{1}{4}r^4,
    \nonumber\\
    K_{\mathrm{Sch}}^{\mathrm{XY}}
    &=
    2h_2(s^2).
    \label{eq:appA-xy-quantities}
\end{align}
The maximum product-input concurrence is
\begin{equation}
    C_{\max}^{\mathrm{XY}}
    =
    \begin{cases}
        \sin(4\theta),
        &
        0\leq\theta<\pi/8,
        \\[1mm]
        1,
        &
        \pi/8\leq\theta\leq\pi/4.
    \end{cases}
    \label{eq:appA-xy-cmax}
\end{equation}
Thus $\sqrt{\mathrm{iSWAP}}$, at $\theta=\pi/8$, is where this
family first enters the perfect-entangler region.

\subsection{Fractional SWAP family}
\label{app:family-fswap}

For the fractional SWAP family $(\theta,\theta,\theta)$, the
operator-Schmidt spectrum has the $1+3$ form
\begin{equation}
    \mathbf p_{\mathrm{fSWAP}}
    =
    \left(
        1-\frac{3}{4}r^2,\,
        \frac{1}{4}r^2,\,
        \frac{1}{4}r^2,\,
        \frac{1}{4}r^2
    \right).
    \label{eq:appA-fswap-spectrum}
\end{equation}
The corresponding quantities are
\begin{align}
    \gamma_{\mathrm{fSWAP}}
    &=
    2\left(
        \sqrt{1-\frac{3}{4}r^2}
        +\frac{3}{2}r
    \right)^2-1,
    \nonumber\\
    e_{\mathrm p}^{\mathrm{fSWAP}}
    &=
    \frac{2}{3}r^2(1-r^2),
    \nonumber\\
    g_{\mathrm t}^{\mathrm{fSWAP}}
    &=
    r^2,
    \nonumber\\
    E_{\mathrm{op}}^{\mathrm{fSWAP}}
    &=
    \frac{3}{4}r^2(2-r^2),
    \nonumber\\
    K_{\mathrm{Sch}}^{\mathrm{fSWAP}}
    &=
    h_2\!\left(\frac{3r^2}{4}\right)
    +\frac{3r^2}{4}\log_2 3,
    \nonumber\\
    C_{\max}^{\mathrm{fSWAP}}
    &=
    \sin(4\theta)
    =
    2r\sqrt{1-r^2}.
    \label{eq:appA-fswap-quantities}
\end{align}
The spectrum becomes uniform at the SWAP endpoint. At
$\theta=\pi/8$, the family passes through $\sqrt{\mathrm{SWAP}}$,
where both $e_{\mathrm p}$ and $C_{\max}$ attain their maxima along
this edge.

\subsection{CNOT--iSWAP family}
\label{app:family-cnot-iswap}

For the CNOT--iSWAP edge $(\pi/4,\theta,0)$,
\begin{equation}
    \mathbf p_{\mathrm{CNOT-iSWAP}}
    =
    \left(
        \frac{c^2}{2},\,
        \frac{c^2}{2},\,
        \frac{s^2}{2},\,
        \frac{s^2}{2}
    \right).
    \label{eq:appA-cnot-iswap-spectrum}
\end{equation}
The corresponding quantities are
\begin{align}
    \gamma_{\mathrm{CNOT-iSWAP}}
    &=
    3+4r,
    \nonumber\\
    e_{\mathrm p}^{\mathrm{CNOT-iSWAP}}
    &=
    \frac{2}{9},
    \nonumber\\
    g_{\mathrm t}^{\mathrm{CNOT-iSWAP}}
    &=
    \frac{1}{3}(1+r^2),
    \nonumber\\
    E_{\mathrm{op}}^{\mathrm{CNOT-iSWAP}}
    &=
    \frac{1}{2}+\frac{1}{4}r^2,
    \nonumber\\
    K_{\mathrm{Sch}}^{\mathrm{CNOT-iSWAP}}
    &=
    1+h_2(s^2),
    \nonumber\\
    C_{\max}^{\mathrm{CNOT-iSWAP}}
    &=
    1.
    \label{eq:appA-cnot-iswap-quantities}
\end{align}
Thus both $e_{\mathrm p}$ and $C_{\max}$ are constant while
$\gamma$ increases from $3$ to $7$ along this edge.

\subsection{iSWAP--SWAP family}
\label{app:family-iswap-swap}

For the iSWAP--SWAP edge $(\pi/4,\pi/4,\theta)$, all four
operator-Schmidt probabilities are equal:
\begin{equation}
    \mathbf p_{\mathrm{iSWAP-SWAP}}
    =
    \left(
        \frac{1}{4},\,
        \frac{1}{4},\,
        \frac{1}{4},\,
        \frac{1}{4}
    \right).
    \label{eq:appA-iswap-swap-spectrum}
\end{equation}
Consequently,
\begin{align}
    \gamma_{\mathrm{iSWAP-SWAP}}
    &=
    7,
    \nonumber\\
    e_{\mathrm p}^{\mathrm{iSWAP-SWAP}}
    &=
    \frac{2}{9}(1-r^2),
    \nonumber\\
    g_{\mathrm t}^{\mathrm{iSWAP-SWAP}}
    &=
    \frac{1}{3}(2+r^2),
    \nonumber\\
    E_{\mathrm{op}}^{\mathrm{iSWAP-SWAP}}
    &=
    \frac{3}{4},
    \nonumber\\
    K_{\mathrm{Sch}}^{\mathrm{iSWAP-SWAP}}
    &=
    2,
    \nonumber\\
    C_{\max}^{\mathrm{iSWAP-SWAP}}
    &=
    \sqrt{1-r^2}.
    \label{eq:appA-iswap-swap-quantities}
\end{align}
In particular,
\begin{equation}
    e_{\mathrm p}^{\mathrm{iSWAP-SWAP}}
    =
    \frac{2}{3}
    \left(
        1-g_{\mathrm t}^{\mathrm{iSWAP-SWAP}}
    \right)
    =
    \frac{2}{9}
    \left(
        C_{\max}^{\mathrm{iSWAP-SWAP}}
    \right)^2.
    \label{eq:appA-iswap-swap-relations}
\end{equation}

\subsection{CNOT--SWAP family}
\label{app:family-cnot-swap}

For the CNOT--SWAP edge $(\pi/4,\theta,\theta)$, the spectrum has
the pair-pair form
\begin{equation}
    \mathbf p_{\mathrm{CNOT-SWAP}}
    =
    \left(
        \frac{2-r^2}{4},\,
        \frac{2-r^2}{4},\,
        \frac{r^2}{4},\,
        \frac{r^2}{4}
    \right).
    \label{eq:appA-cnot-swap-spectrum}
\end{equation}
The corresponding quantities are
\begin{align}
    \gamma_{\mathrm{CNOT-SWAP}}
    &=
    3+4r\sqrt{2-r^2},
    \nonumber\\
    e_{\mathrm p}^{\mathrm{CNOT-SWAP}}
    &=
    \frac{2}{9}(1-r^4),
    \nonumber\\
    g_{\mathrm t}^{\mathrm{CNOT-SWAP}}
    &=
    \frac{1}{3}(1+2r^2),
    \nonumber\\
    E_{\mathrm{op}}^{\mathrm{CNOT-SWAP}}
    &=
    \frac{1}{2}
    +\frac{1}{2}r^2
    -\frac{1}{4}r^4,
    \nonumber\\
    K_{\mathrm{Sch}}^{\mathrm{CNOT-SWAP}}
    &=
    1+h_2\!\left(\frac{r^2}{2}\right).
    \label{eq:appA-cnot-swap-quantities}
\end{align}
Its maximum product-input concurrence is
\begin{equation}
    C_{\max}^{\mathrm{CNOT-SWAP}}
    =
    \begin{cases}
        1,
        &
        0\leq\theta\leq\pi/8,
        \\[1mm]
        \sin(4\theta),
        &
        \pi/8<\theta\leq\pi/4.
    \end{cases}
    \label{eq:appA-cnot-swap-cmax}
\end{equation}
Hence this family leaves the perfect-entangler region at
$\theta=\pi/8$. Its \gls{QPD} extent takes the endpoint values
$\gamma=3$ at CNOT and $\gamma=7$ at SWAP.

\section{Proof of the Gate-Action Envelopes}
\label{app:gate-action-extremality}

This appendix proves Theorem~\ref{thm:gate-action-envelopes}. The
entangling-power bounds follow from two global inequalities. For gate
typicality, we reduce the problem to a symmetric optimisation on a
cube slice, classify its boundary and interior stationary points, and
then compare the resulting canonical branches.

\subsection{Entangling-power bounds}
\label{app:ep-proof}

The controlled family satisfies
$e_{\mathrm p}=2r^2/9$ and $\gamma=1+2r$ by
Appendix~\ref{app:family-controlled}. Eliminating $r$ gives
\begin{equation}
    \gamma
    =
    1+3\sqrt{2e_{\mathrm p}},
    \label{eq:appB-ep-controlled}
\end{equation}
which supplies the candidate lower boundary.

To prove globality, Eq.~\eqref{eq:gamma-schmidt} gives
\begin{align}
    \gamma-1
    &=
    4\sum_{j<k}\sqrt{p_jp_k}
    \nonumber\\
    &\geq
    4\sqrt{\sum_{j<k}p_jp_k}
    \nonumber\\
    &=
    2\sqrt{2E_{\mathrm{op}}},
    \label{eq:gamma-eop-sqrt-bound}
\end{align}
where the inequality follows from non-negativity and the final equality
uses
$E_{\mathrm{op}}=2\sum_{j<k}p_jp_k$.

Next define
\begin{equation}
    x_j=\cos(4\theta_j).
    \label{eq:appB-xj-definition}
\end{equation}
Equations~\eqref{eq:ep-cartan} and \eqref{eq:gt-cartan} imply
\begin{align}
    g_{\mathrm t}-\frac32e_{\mathrm p}
    &=
    \frac{1}{12}
    \big[
        (1-x_1)(1-x_2)
        +(1-x_2)(1-x_3)
        +(1-x_3)(1-x_1)
    \big]
    \nonumber\\
    &\geq0,
    \label{eq:gt-ep-lower}
\end{align}
because $-1\leq x_j\leq1$. Combining
Eq.~\eqref{eq:gt-ep-lower} with Eq.~\eqref{eq:eop-ep-gt} gives
\begin{equation}
    E_{\mathrm{op}}
    \geq
    \frac94e_{\mathrm p}.
    \label{eq:eop-ep-lower}
\end{equation}
Substitution into Eq.~\eqref{eq:gamma-eop-sqrt-bound} yields
\begin{equation}
    \gamma
    \geq
    1+3\sqrt{2e_{\mathrm p}}.
    \label{eq:ep-lower-proof}
\end{equation}

The equality conditions identify the lower family. Equality in the
first inequality of Eq.~\eqref{eq:gamma-eop-sqrt-bound} requires at
most one product $\sqrt{p_jp_k}$ to be nonzero, so the
operator-Schmidt spectrum has rank at most two. Equality in
Eq.~\eqref{eq:gt-ep-lower} requires at most one of the quantities
$1-x_j$ to be nonzero. In the non-negative representative tetrahedron
this gives $\theta_2=\theta_3=0$, namely the controlled family
(including the identity endpoint).

For the upper bound, Cauchy--Schwarz gives
\begin{equation}
    \sum_{j=0}^{3}\sqrt{p_j}
    \leq
    \sqrt{4\sum_{j=0}^{3}p_j}
    =
    2,
    \label{eq:appB-schmidt-cauchy}
\end{equation}
and hence
\begin{equation}
    \gamma\leq7.
    \label{eq:gamma-global-upper}
\end{equation}
Equality requires $p_j=1/4$ for every $j$. The iSWAP--SWAP family has
this uniform spectrum throughout and
$e_{\mathrm p}=2(1-r^2)/9$, so it attains $\gamma=7$ for every
$e_{\mathrm p}\in[0,2/9]$. This proves
Eq.~\eqref{eq:ep-envelope}.

\subsection{Gate-typicality reduction}
\label{app:gt-reduction}

Define
\begin{equation}
    a=\sin^2(2\theta_1),
    \qquad
    b=\sin^2(2\theta_2),
    \qquad
    c=\sin^2(2\theta_3),
    \qquad
    Q=3g_{\mathrm t}.
    \label{eq:appB-abc}
\end{equation}
Then
\begin{equation}
    a+b+c=Q,
    \qquad
    0\leq a,b,c\leq1.
    \label{eq:appB-cube-slice}
\end{equation}
In the non-negative tetrahedron $a\geq b\geq c$, but the objective
below is symmetric under permutations of $a,b,c$. Consequently, the
extrema over the ordered physical domain are the same as the extrema
over the complete cube slice
$[0,1]^3\cap\{a+b+c=Q\}$.

Let
\begin{equation}
    t_j=\cos(2\theta_j),
    \qquad
    t_1^2=1-a,
    \quad
    t_2^2=1-b,
    \quad
    t_3^2=1-c.
    \label{eq:appB-tj}
\end{equation}
The operator-Schmidt probabilities can then be written as
\begin{align}
    p_0
    &=
    \frac{1+t_1t_2+t_1t_3+t_2t_3}{4},
    \nonumber\\
    p_1
    &=
    \frac{1+t_1t_2-t_1t_3-t_2t_3}{4},
    \nonumber\\
    p_2
    &=
    \frac{1-t_1t_2+t_1t_3-t_2t_3}{4},
    \nonumber\\
    p_3
    &=
    \frac{1-t_1t_2-t_1t_3+t_2t_3}{4}.
    \label{eq:appB-p-t}
\end{align}
Define
\begin{equation}
    X_1
    =
    2\left(\sqrt{p_0p_1}+\sqrt{p_2p_3}\right),
    \label{eq:appB-X1}
\end{equation}
with $X_2$ and $X_3$ obtained cyclically. Equation~\eqref{eq:gamma-schmidt}
becomes
\begin{equation}
    \gamma
    =
    1+2(X_1+X_2+X_3).
    \label{eq:appB-gamma-X}
\end{equation}
Introduce
\begin{equation}
    \Delta
    =
    (ab+ac+bc)^2
    -
    4abc(a+b+c-1).
    \label{eq:appB-Delta}
\end{equation}
Direct simplification gives
\begin{equation}
    X_1^2
    =
    \frac{
        2a-ab-ac+bc+\sqrt{\Delta}
    }{2},
    \label{eq:appB-X1-squared}
\end{equation}
with cyclic expressions for $X_2^2$ and $X_3^2$. The fixed-$g_{\mathrm t}$
problem is therefore a compact symmetric constrained optimisation of
$X_1+X_2+X_3$ on Eq.~\eqref{eq:appB-cube-slice}.

\subsection{Boundary faces and canonical branches}
\label{app:gt-faces}

By permutation symmetry, it is sufficient to analyse one zero face and
one unit face.

\subsubsection{The face \texorpdfstring{$c=0$}{c=0}}

Setting $c=0$ in Eq.~\eqref{eq:appB-Delta} gives
$\sqrt{\Delta}=ab$, and hence
\begin{equation}
    X_1=\sqrt a,
    \qquad
    X_2=\sqrt b,
    \qquad
    X_3=\sqrt{ab}.
    \label{eq:appB-c0-X}
\end{equation}
Thus
\begin{equation}
    \gamma
    =
    1+2\left(\sqrt a+\sqrt b+\sqrt{ab}\right),
    \qquad
    a+b=Q.
    \label{eq:appB-c0-gamma}
\end{equation}
After substituting $b=Q-a$, direct differentiation gives a negative
second derivative throughout the interior of the allowed line segment.
The objective is therefore strictly concave and, by symmetry, its
maximum is at $a=b=Q/2$, giving the XY/iSWAP branch
\begin{equation}
    \gamma_{\mathrm{XY}}
    =
    1+Q+4\sqrt{\frac Q2}
    =
    1+3g_{\mathrm t}
    +4\sqrt{\frac{3g_{\mathrm t}}{2}}.
    \label{eq:appB-gamma-XY}
\end{equation}
The minimum of a concave function on this segment occurs at an endpoint.
For $0\leq Q\leq1$ the endpoint $(Q,0,0)$ gives the controlled branch
\begin{equation}
    \gamma_{\mathrm C}
    =
    1+2\sqrt Q
    =
    1+2\sqrt{3g_{\mathrm t}}.
    \label{eq:appB-gamma-C}
\end{equation}
For $1\leq Q\leq2$, the endpoint $(1,Q-1,0)$ gives the auxiliary
CNOT--iSWAP branch
\begin{equation}
    \gamma_{\mathrm{CI}}
    =
    3+4\sqrt{Q-1}
    =
    3+4\sqrt{3g_{\mathrm t}-1}.
    \label{eq:appB-gamma-CI}
\end{equation}

\subsubsection{The face \texorpdfstring{$a=1$}{a=1}}

For $a=1$,
\begin{equation}
    \sqrt{\Delta}=b+c-bc,
    \qquad
    X_1=1,
    \qquad
    X_2=X_3=\sqrt{b+c-bc},
    \label{eq:appB-a1-X}
\end{equation}
so
\begin{equation}
    \gamma
    =
    3+4\sqrt{b+c-bc}.
    \label{eq:appB-a1-gamma}
\end{equation}
Let $s=b+c=Q-1$. At fixed $s$, minimising $\gamma$ is equivalent to
maximising $bc$, whose maximum occurs at $b=c=s/2$. This gives the
CNOT--SWAP branch
\begin{equation}
    \gamma_{\mathrm{CS}}
    =
    3+2\sqrt{(Q-1)(5-Q)}
    =
    3+2\sqrt{(3g_{\mathrm t}-1)(5-3g_{\mathrm t})}.
    \label{eq:appB-gamma-CS}
\end{equation}
To maximise Eq.~\eqref{eq:appB-a1-gamma}, $bc$ must instead be
minimised. For $1\leq Q\leq2$, its minimum is zero and reproduces the
CNOT--iSWAP branch. For $2\leq Q\leq3$, the constraints
$0\leq b,c\leq1$ imply the minimum $bc=Q-2$, attained when one of
$b,c$ equals one. Equation~\eqref{eq:appB-a1-gamma} then gives
\begin{equation}
    \gamma=7,
    \qquad
    2\leq Q\leq3,
    \label{eq:appB-gamma-plateau}
\end{equation}
which is the iSWAP--SWAP plateau.

\subsubsection{The symmetric interior point}

At the symmetric point
\begin{equation}
    a=b=c=\frac Q3=g_{\mathrm t},
    \label{eq:appB-symmetric-point}
\end{equation}
the Cartan parameters lie on the fractional SWAP family and
\begin{equation}
    \gamma_{\mathrm{FS}}
    =
    1+3g_{\mathrm t}
    +
    3\sqrt{g_{\mathrm t}(4-3g_{\mathrm t})}.
    \label{eq:appB-gamma-FS}
\end{equation}

\subsection{Interior stationary points}
\label{app:gt-interior}

It remains to show that no additional interior stationary point can
generate a global boundary.

\subsubsection{Two-equal nonsymmetric stationary points}

Suppose
\begin{equation}
    b=c=y,
    \qquad
    a=x,
    \qquad
    Q=x+2y,
    \label{eq:appB-two-equal}
\end{equation}
and define
\begin{equation}
    W=\sqrt{y^2+4x(1-y)}.
    \label{eq:appB-W}
\end{equation}
Then
\begin{equation}
    \frac{\gamma-1}{4}
    =
    \frac{W+y}{4}
    +
    \sqrt{
        \frac{y(2-y+W)}{2}
    }.
    \label{eq:appB-two-equal-objective}
\end{equation}
Substituting $x=Q-2y$, differentiating at fixed $Q$, and eliminating
the radicals gives the stationary resultant
\begin{equation}
    (Q-3y)^2(Q-1)^2(y-1)
    \left[
        Qy^2+18y^3-61y^2+56y-16
    \right]
    =
    0.
    \label{eq:appB-two-equal-resultant}
\end{equation}
The factor $Q-3y$ is the symmetric solution. The factors $Q-1$ and
$y-1$ are boundary or radical-elimination degeneracies and do not
satisfy the unsquared interior stationarity equation as additional
branches. The genuine nonsymmetric branch is therefore
\begin{align}
    W
    &=
    \frac{(3y-4)(3y-2)}{y},
    \nonumber\\
    Q
    &=
    -\frac{18y^3-61y^2+56y-16}{y^2},
    \nonumber\\
    x
    &=
    -\frac{(5y-4)(4y^2-9y+4)}{y^2}.
    \label{eq:appB-two-equal-branch}
\end{align}
For two independent tangent directions to the fixed-$Q$ plane, the
second derivatives are proportional to
\begin{align}
    L
    &=
    -
    \frac{
        (y-2)^2(y+2)(7y-4)
    }{
        8y^2(y-1)(3y-4)(3y-2)
    },
    \nonumber\\
    T
    &=
    -
    \frac{
        2(y-2)^2(y-1)(7y-4)
    }{
        y^4(3y-4)
    }.
    \label{eq:appB-LT}
\end{align}
Their product is
\begin{equation}
    LT
    =
    \frac{
        (y-2)^4(y+2)(7y-4)^2
    }{
        4y^6(3y-4)^2(3y-2)
    }.
    \label{eq:appB-LT-product}
\end{equation}
On an interior physical branch $0<y<1$, so $3y-4<0$, while
$W>0$. The first relation in Eq.~\eqref{eq:appB-two-equal-branch}
therefore forces $3y-2<0$, or $y<2/3$. It follows from
Eq.~\eqref{eq:appB-LT-product} that $LT<0$. Every such stationary
point is therefore a saddle. The zero at $y=4/7$ coincides with the
symmetric fractional SWAP point.

\subsubsection{Fully asymmetric stationary points}

Let
\begin{equation}
    F(a,b,c)=X_1+X_2+X_3
    \label{eq:appB-F}
\end{equation}
and introduce the elementary symmetric variables
\begin{equation}
    s=a+b+c,
    \qquad
    q=ab+ac+bc,
    \qquad
    \rho=abc.
    \label{eq:appB-sqrho}
\end{equation}
Because $F$ is symmetric, write locally $F=\Phi(s,q,\rho)$. Define
\begin{equation}
    R=\sqrt{q^2-4\rho(s-1)}
    \label{eq:appB-R}
\end{equation}
and $y_i=X_i^2$. Their symmetric combinations are
\begin{align}
    A:=\sum_i y_i
    &=
    \frac{2s-q+3R}{2},
    \nonumber\\
    B:=\sum_{i<j}y_i y_j
    &=
    \frac{
        R(2s-q)+q^2+2q-4\rho s
    }{2},
    \nonumber\\
    C:=y_1y_2y_3
    &=
    \frac{
        Rq-2R\rho+q^2-2q\rho
        +2\rho^2-2\rho s+2\rho
    }{2}.
    \label{eq:appB-ABC}
\end{align}
Let
\begin{equation}
    P=X_1X_2+X_1X_3+X_2X_3,
    \qquad
    Z=X_1X_2X_3=\sqrt C.
    \label{eq:appB-PZ}
\end{equation}
Using $F^2=A+2P$ and $P^2=B+2FZ$ gives
\begin{equation}
    (F^2-A)^2-4B-8F\sqrt C=0.
    \label{eq:appB-F-implicit}
\end{equation}
Differentiating at fixed $s$ and $\rho$ yields
\begin{equation}
    \Phi_q
    =
    \frac{
        P A_q+B_q+F C_q/Z
    }{
        2(FP-Z)
    },
    \label{eq:appB-Phiq}
\end{equation}
whose denominator is strictly positive in the physical interior:
for $X_i>0$, expansion of $FP$ gives $FP-Z>0$. The required
derivatives are
\begin{align}
    A_q
    &=
    \frac{3q-R}{2R},
    \nonumber\\
    B_q
    &=
    1+\frac{2qs-(R-q)^2}{2R},
    \nonumber\\
    C_q
    &=
    \frac{(R+q)(R+q-2\rho)}{2R}.
    \label{eq:appB-ABCq}
\end{align}

We now establish $\Phi_q>0$. First, $B_q>1$. If $s\geq1$, then
$R\leq q$ and $q\leq s^2/3\leq s$, which gives
$(R-q)^2<2qs$. If $0<s<1$, the inequalities
$q^2\geq3s\rho$ and $q\leq s^2/3$ give
\begin{equation}
    R^2
    \leq
    q^2\frac{4-s}{3s}
    \leq
    qs\frac{4-s}{9}
    <
    2qs,
    \label{eq:appB-R-bound}
\end{equation}
and again $(R-q)^2<2qs$. Second, the arithmetic--geometric mean inequality gives
$q\geq3\rho^{2/3}>2\rho$ in the interior, hence $C_q>0$.

If $A_q\geq0$, Eq.~\eqref{eq:appB-Phiq} is therefore positive
immediately. Suppose instead that $A_q<0$, equivalently $R>3q$.
This cannot occur for $s\geq1$, so $s<1$. Moreover,
$R^2=q^2+4\rho(1-s)>9q^2$ implies
$\rho(1-s)>2q^2$. Combining this with $q^2\geq3s\rho$ yields
$s<1/7$. In this corner,
\begin{equation}
    q\leq\frac{s^2}{3},
    \qquad
    \rho\leq\frac{s^3}{27},
    \qquad
    R<s.
    \label{eq:appB-small-s}
\end{equation}
Consequently
\begin{equation}
    A
    =
    \frac{2s-q+3R}{2}
    <
    \frac52s
    <
    1.
    \label{eq:appB-A-small}
\end{equation}
Since $2X_iX_j\leq X_i^2+X_j^2$, we have $P\leq A<1$, while
$A_q>-1/2$ and $B_q>1$. Thus
$P A_q+B_q>1/2$, and the remaining term $F C_q/Z$ is strictly
positive. Hence
\begin{equation}
    \Phi_q>0
    \label{eq:appB-Phiq-positive}
\end{equation}
throughout the physical interior.

At a fully asymmetric constrained stationary point, $a,b,c$ are
pairwise distinct. At fixed $s$,
\begin{equation}
    F_a-F_b
    =
    (b-a)(\Phi_q+c\Phi_\rho),
    \label{eq:appB-Fab}
\end{equation}
with cyclic analogues. Stationarity would require
$\Phi_q+c\Phi_\rho=0$ and $\Phi_q+b\Phi_\rho=0$. Subtracting gives
$(c-b)\Phi_\rho=0$, so $\Phi_\rho=0$ and then $\Phi_q=0$, contrary
to Eq.~\eqref{eq:appB-Phiq-positive}. There are therefore no fully
asymmetric interior stationary extrema.

It follows that the only interior extremum that can contribute to a
global envelope is the symmetric fractional SWAP point; every other
global extremum lies on the boundary faces already classified.

\subsection{Branch comparisons and completion}
\label{app:gt-completion}

For compactness write $g=g_{\mathrm t}$. The four nonconstant
candidate branches are
\begin{align}
    \gamma_{\mathrm C}(g)
    &=
    1+2\sqrt{3g},
    \nonumber\\
    \gamma_{\mathrm{CS}}(g)
    &=
    3+2\sqrt{(3g-1)(5-3g)},
    \nonumber\\
    \gamma_{\mathrm{FS}}(g)
    &=
    1+3g+3\sqrt{g(4-3g)},
    \nonumber\\
    \gamma_{\mathrm{XY}}(g)
    &=
    1+3g+4\sqrt{\frac{3g}{2}}.
    \label{eq:appB-candidate-branches}
\end{align}

\subsubsection{Upper-envelope ordering}

Equating the fractional SWAP and XY/iSWAP branches gives
\begin{equation}
    3\sqrt{g(4-3g)}
    =
    4\sqrt{\frac{3g}{2}}.
    \label{eq:gt-upper-crossing}
\end{equation}
Besides the common identity endpoint $g=0$, division by $\sqrt g$
and squaring gives
\begin{equation}
    g=\frac49.
    \label{eq:gt-crossing-four-ninths}
\end{equation}
Substitution gives the crossing value in
Eq.~\eqref{eq:gt-four-ninths-gamma}. Direct comparison gives
$\gamma_{\mathrm{FS}}>\gamma_{\mathrm{XY}}$ for
$0<g<4/9$ and
$\gamma_{\mathrm{XY}}>\gamma_{\mathrm{FS}}$ for
$4/9<g\leq2/3$. At $g=2/3$, the XY/iSWAP branch reaches
$\gamma=7$ and joins the iSWAP--SWAP plateau. Since
Eq.~\eqref{eq:gamma-global-upper} is universal, this plateau is the
global upper envelope for $2/3\leq g\leq1$.

\subsubsection{Lower-envelope ordering}

At the CNOT class,
\begin{equation}
    g=\frac13,
    \qquad
    \gamma_{\mathrm C}
    =
    \gamma_{\mathrm{CS}}
    =
    3.
    \label{eq:appB-lower-CNOT}
\end{equation}
For $0<g\leq1/3$, $4-3g\geq3$ and hence
\begin{equation}
    3\sqrt{g(4-3g)}
    >
    2\sqrt{3g},
    \label{eq:appB-C-vs-FS}
\end{equation}
so the fractional SWAP branch cannot undercut the controlled branch.

Equating the CNOT--SWAP and fractional SWAP branches and eliminating
the square roots gives
\begin{equation}
    (g-1)^2(3g-2)^2=0.
    \label{eq:appB-CS-FS-crossings}
\end{equation}
Direct substitution verifies the two intersections on their common
domain, $g=2/3$ and the common SWAP endpoint $g=1$. A sign check on
the intervals between them gives
\begin{equation}
    \gamma_{\mathrm{CS}}<\gamma_{\mathrm{FS}}
    \quad
    \left(\frac13\leq g<\frac23\right),
    \qquad
    \gamma_{\mathrm{FS}}<\gamma_{\mathrm{CS}}
    \quad
    \left(\frac23<g<1\right).
    \label{eq:appB-CS-FS-order}
\end{equation}
Thus the lower envelope changes from controlled to CNOT--SWAP at
$g=1/3$ and from CNOT--SWAP to fractional SWAP at $g=2/3$.

\subsubsection{Auxiliary CNOT--iSWAP branch}

It remains to exclude the auxiliary branch
\begin{equation}
    \gamma_{\mathrm{CI}}
    =
    3+4\sqrt{Q-1},
    \qquad
    1\leq Q\leq2.
    \label{eq:appB-CI-again}
\end{equation}
For the lower envelope,
\begin{equation}
    \gamma_{\mathrm{CS}}
    =
    3+2\sqrt{(Q-1)(5-Q)}
    \leq
    3+4\sqrt{Q-1}
    =
    \gamma_{\mathrm{CI}},
    \label{eq:appB-CI-lower}
\end{equation}
because $\sqrt{5-Q}\leq2$ on this interval.

For $1\leq Q\leq4/3$, the relevant upper branch is fractional SWAP.
Set $x=\sqrt{Q-1}$, so $0\leq x\leq1/\sqrt3$. After squaring
positive quantities, $\gamma_{\mathrm{CI}}\leq\gamma_{\mathrm{FS}}$
is equivalent to
\begin{equation}
    f(x)
    =
    x^4-2x^3+2x^2+2x-2
    \leq0.
    \label{eq:appB-fx}
\end{equation}
Here
\begin{equation}
    f'(x)
    =
    2+2x(2x^2-3x+2)>0,
    \label{eq:appB-fprime}
\end{equation}
and
\begin{equation}
    f\left(\frac1{\sqrt3}\right)
    =
    \frac{-11+4\sqrt3}{9}
    <
    0.
    \label{eq:appB-fendpoint}
\end{equation}
Hence $\gamma_{\mathrm{CI}}<\gamma_{\mathrm{FS}}$ throughout this
range. For $4/3\leq Q\leq2$, the relevant upper branch is XY/iSWAP,
and the same substitution reduces
$\gamma_{\mathrm{CI}}\leq\gamma_{\mathrm{XY}}$ to
\begin{equation}
    (7-x)(1-x)^2(1+x)\geq0,
    \qquad
    \frac1{\sqrt3}\leq x\leq1.
    \label{eq:appB-CI-XY}
\end{equation}
This is immediate, with equality only at $x=1$, corresponding to
iSWAP. The auxiliary CNOT--iSWAP branch therefore lies between the
final lower and upper envelopes throughout its domain.

The boundary classification, interior exclusion and branch comparisons
are exhaustive. They reproduce Eqs.~\eqref{eq:gt-lower-envelope} and
\eqref{eq:gt-upper-envelope}. Up to permutations and the local-equivalence
symmetries already removed by the Cartan convention, the equality families
are exactly those stated in Theorem~\ref{thm:gate-action-envelopes}; at
the transition values, the adjacent equality families meet or coexist.
\section{Physical Operator-Schmidt Spectra}
\label{app:physical-schmidt-spectra}

This appendix derives the physical spectrum condition stated in
Lemma~\ref{lem:physical-schmidt-spectrum}, its rank consequences, and
the factorisation and attainability of the physical boundary. The
rank-three exclusion itself is known from earlier work
\cite{Dur_2002,Tyson_2003}; here we derive the ordered
necessary-and-sufficient condition used in the subsequent extremal
analysis. The starting point is the one-qubit unistochastic
characterisation of Musz, Ku\'s and \.{Z}yczkowski
\cite{Musz_2013}, together with their identification of the spectrum
of the associated dynamical matrix with the normalised
operator-Schmidt spectrum of the underlying two-qubit unitary.

\subsection{Ordered-spectrum condition}
\label{app:physical-spectrum-condition}

Let the normalised operator-Schmidt probabilities be ordered as in
Eq.~\eqref{eq:ordered-physical-spectrum}. In the unistochastic
description, introduce the damping coordinates
\begin{align}
    \eta_1
    &=
    p_{(1)}+p_{(2)}-p_{(3)}-p_{(4)},
    \nonumber\\
    \eta_2
    &=
    p_{(1)}-p_{(2)}+p_{(3)}-p_{(4)},
    \nonumber\\
    \eta_3
    &=
    p_{(1)}-p_{(2)}-p_{(3)}+p_{(4)}.
    \label{eq:appC-eta-from-p}
\end{align}
For one-qubit bistochastic maps, the necessary and sufficient
unistochastic conditions can be written \cite{Musz_2013} as
\begin{equation}
    \eta_1\eta_2\leq\eta_3,
    \qquad
    \eta_2\eta_3\leq\eta_1,
    \qquad
    \eta_3\eta_1\leq\eta_2.
    \label{eq:appC-unistochastic-eta}
\end{equation}
Using $\sum_j p_{(j)}=1$, these inequalities factor respectively as
\begin{align}
    p_{(1)}p_{(4)}
    &\geq
    p_{(2)}p_{(3)},
    \nonumber\\
    p_{(1)}p_{(2)}
    &\geq
    p_{(3)}p_{(4)},
    \nonumber\\
    p_{(1)}p_{(3)}
    &\geq
    p_{(2)}p_{(4)}.
    \label{eq:appC-factorised-conditions}
\end{align}
After ordering, the final two inequalities are automatic:
$p_{(1)}\geq p_{(3)}$ and $p_{(2)}\geq p_{(4)}$ imply the second,
while $p_{(1)}\geq p_{(2)}$ and
$p_{(3)}\geq p_{(4)}$ imply the third. The only nontrivial condition
is therefore
\begin{equation}
    p_{(1)}p_{(4)}
    \geq
    p_{(2)}p_{(3)}.
    \label{eq:appC-physical-condition}
\end{equation}
Within the unistochastic correspondence, Eq.~\eqref{eq:appC-physical-condition}
is thus both necessary and sufficient for an ordered normalised
four-component spectrum to be compatible with a two-qubit unitary.

\subsection{Rank consequences}
\label{app:physical-spectrum-ranks}

The condition immediately recovers the known exclusion of operator-Schmidt rank three. If
$p_{(4)}=0$, Eq.~\eqref{eq:appC-physical-condition} gives
\begin{equation}
    p_{(2)}p_{(3)}=0.
    \label{eq:appC-rank3-product}
\end{equation}
Since $p_{(2)}\geq p_{(3)}\geq0$, this forces
$p_{(3)}=0$. Hence a two-qubit unitary has operator-Schmidt rank one,
two or four, but never three.

Rank-one spectra have the form
\begin{equation}
    \mathbf p=(1,0,0,0),
    \label{eq:appC-rank1}
\end{equation}
and correspond to local gates up to local equivalence. Ordered
rank-two spectra have the form
\begin{equation}
    \mathbf p=(x,1-x,0,0),
    \qquad
    \frac12\leq x\leq1.
    \label{eq:appC-rank2}
\end{equation}
They are all attained by the controlled family, for which
$x=\cos^2\theta$ in Eq.~\eqref{eq:appA-controlled-spectrum}.
All remaining nontrivial physical spectra have rank four.

\subsection{Factorisation of the physical boundary}
\label{app:physical-spectrum-boundary}

For rank-four spectra, the nontrivial boundary obtained by saturating
the unistochastic constraint is
\begin{equation}
    p_{(1)}p_{(4)}
    =
    p_{(2)}p_{(3)}.
    \label{eq:appC-boundary-equality}
\end{equation}
Equivalently,
\begin{equation}
    \det
    \begin{pmatrix}
        p_{(1)} & p_{(2)}
        \\
        p_{(3)} & p_{(4)}
    \end{pmatrix}
    =
    0.
    \label{eq:appC-boundary-determinant}
\end{equation}
The non-negative matrix in Eq.~\eqref{eq:appC-boundary-determinant}
therefore has rank one. After a possible permutation of the four
entries, every boundary spectrum factorises as
\begin{equation}
    \mathbf p
    =
    (x,1-x)\otimes(y,1-y),
    \qquad
    0\leq x,y\leq1.
    \label{eq:appC-boundary-factorisation}
\end{equation}
Equivalently,
\begin{equation}
    \mathbf p
    =
    \big(
        xy,\,
        x(1-y),\,
        (1-x)y,\,
        (1-x)(1-y)
    \big)
    \label{eq:appC-boundary-components}
\end{equation}
up to permutation. By exchanging the outcomes of either binary factor,
one may take $x,y\in[1/2,1]$ without loss.

This factorisation is the reason the lower-envelope problem in
Appendix~\ref{app:entropy-envelope-derivations} reduces to an entropy
allocation problem between two binary spectra: for every R\'enyi
order,
\begin{equation}
    H_\alpha(\mathbf p)
    =
    H_\alpha(x,1-x)
    +
    H_\alpha(y,1-y).
    \label{eq:appC-renyi-additivity}
\end{equation}

\subsection{Constructive attainability}
\label{app:physical-spectrum-attainability}

The factorised boundary is not merely an abstract probability-space
constraint. Setting $\theta_3=0$ in
Eq.~\eqref{eq:schmidt-probabilities-cartan} gives, up to permutation,
\begin{equation}
    \mathbf p(\theta_1,\theta_2,0)
    =
    \big(
        \cos^2\theta_1,\,
        \sin^2\theta_1
    \big)
    \otimes
    \big(
        \cos^2\theta_2,\,
        \sin^2\theta_2
    \big).
    \label{eq:appC-theta3-zero-factorisation}
\end{equation}
Conversely, for any $x,y\in[1/2,1]$, choose
\begin{equation}
    \theta_x
    =
    \arccos\sqrt{x},
    \qquad
    \theta_y
    =
    \arccos\sqrt{y},
    \label{eq:appC-angle-construction}
\end{equation}
which lie in $[0,\pi/4]$. Exchanging $x$ and $y$ if necessary places
the two angles in Weyl order. Thus every factorised boundary spectrum
is realised by a physical Cartan point on the $\theta_3=0$ face.

Several canonical spectra used in Theorem~\ref{thm:spectral-envelopes}
appear as special cases. Taking one binary factor to be deterministic
gives the controlled rank-two spectra. Taking one factor to be uniform
gives the pair-pair family
\begin{equation}
    \mathbf p_{\mathrm{pair}}
    =
    \left(
        \frac{1-t}{2},\,
        \frac{1-t}{2},\,
        \frac{t}{2},\,
        \frac{t}{2}
    \right),
    \qquad
    0\leq t\leq\frac12.
    \label{eq:appC-pair-pair}
\end{equation}
On the $\theta_3=0$ face this is realised by the CNOT--iSWAP edge.
Appendix~\ref{app:family-cnot-swap} shows that the CNOT--SWAP family
realises the same one-parameter set of pair-pair spectra. Consequently,
the spectral quantities $\gamma$, $E_{\mathrm{op}}$ and
$K_{\mathrm{Sch}}$ cannot distinguish these two Cartan families along
the corresponding spectrum locus. In Theorem~\ref{thm:spectral-envelopes}
we use CNOT--SWAP as a canonical gate-space representative of this
pair-pair equality spectrum.

At $x=y=1/2$, Eq.~\eqref{eq:appC-boundary-factorisation} becomes the
uniform spectrum
\begin{equation}
    \mathbf p
    =
    \left(
        \frac14,\,
        \frac14,\,
        \frac14,\,
        \frac14
    \right),
    \label{eq:appC-uniform-spectrum}
\end{equation}
attained at iSWAP and, more generally, along the iSWAP--SWAP family
of Appendix~\ref{app:family-iswap-swap}.

For comparison, the fractional SWAP spectrum
\begin{equation}
    \left(
        1-\frac{3x}{4},\,
        \frac{x}{4},\,
        \frac{x}{4},\,
        \frac{x}{4}
    \right)
    \label{eq:appC-fswap-spectrum}
\end{equation}
satisfies
\begin{equation}
    p_{(1)}p_{(4)}-p_{(2)}p_{(3)}
    =
    \frac{x(1-x)}{4}.
    \label{eq:appC-fswap-slack}
\end{equation}
It therefore lies strictly inside the rank-four physical spectrum
region for $0<x<1$, reaching the physical boundary only at its
rank-deficient identity limit and at the uniform SWAP endpoint. This
is consistent with its role as the physical realisation of the
generic upper entropy extremum rather than the sharpened physical
lower boundary.

\begin{figure}[t]
    \centering
    \includegraphics[width=0.82\linewidth]{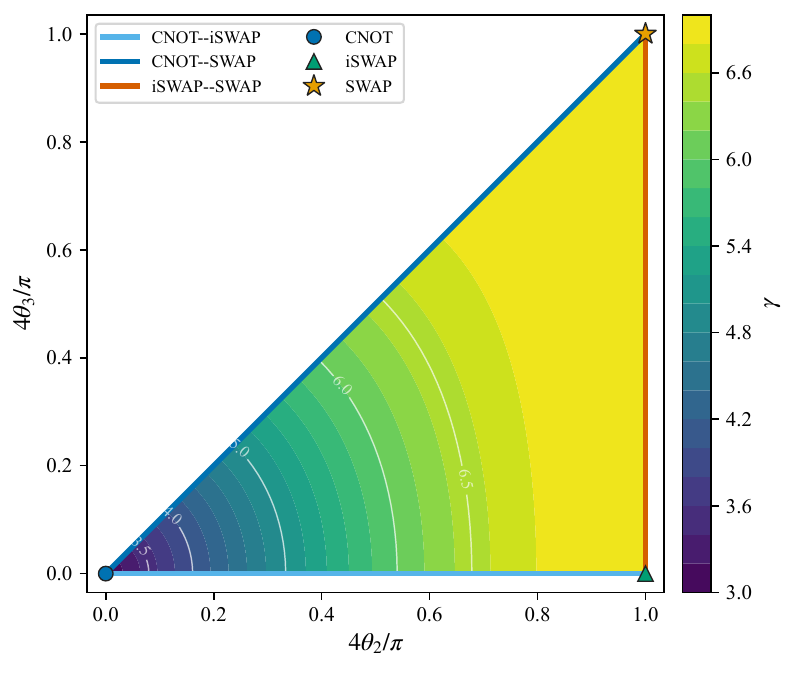}
    \caption{Optimal \gls{QPD} extent on the
    $\theta_1=\pi/4$ face of the non-negative Cartan tetrahedron. This
    face contains the CNOT--iSWAP, CNOT--SWAP and iSWAP--SWAP edges and
    illustrates the physical Cartan geometry associated with the
    pair-pair and uniform operator-Schmidt spectra discussed in this
    appendix.}
    \label{fig:theta1-pi4-spectrum-face}
\end{figure}

Figure~\ref{fig:theta1-pi4-spectrum-face} visualises the
$\theta_1=\pi/4$ Cartan face that contains the pair-pair boundary
families discussed above.

\section{Entropy-Envelope Derivations}
\label{app:entropy-envelope-derivations}

This appendix proves Theorem~\ref{thm:spectral-envelopes}. The upper
spectral envelopes follow by specialising sharp finite-alphabet
entropy-extremality results of Sakai and Iwata
\cite{SakaiIwata_2016_Renyi,SakaiIwata_2016_Shannon} and then showing
that the corresponding extremising probability vectors are physically
realised by two-qubit unitaries. Their papers use natural logarithms
for the entropy definitions; changing the logarithm base only rescales
the entropies and does not change the extremising distributions. The
lower envelopes require the additional physical spectrum restriction
of Lemma~\ref{lem:physical-schmidt-spectrum}.

\subsection{Generic entropy extremality and the upper envelopes}
\label{app:entropy-upper}

Equation~\eqref{eq:gamma-renyi} gives
\begin{equation}
    \gamma
    =
    2^{H_{1/2}(\mathbf p)+1}-1,
    \label{eq:appD-gamma-renyi}
\end{equation}
so maximising $\gamma$ at fixed spectral entropy is equivalent to
maximising $H_{1/2}$. For the upper bounds relevant here, the sharp extremising
distribution of Sakai and Iwata is
\begin{equation}
    \mathbf v_n(y)
    =
    \big(
        1-(n-1)y,\,
        y,\ldots,y
    \big),
    \qquad
    0\leq y\leq\frac{1}{n}.
    \label{eq:appD-vn}
\end{equation}
For four components,
\begin{equation}
    \mathbf v_4(y)
    =
    (1-3y,y,y,y),
    \qquad
    0\leq y\leq\frac14.
    \label{eq:appD-v4}
\end{equation}
Setting $x=4y$ gives
\begin{equation}
    \mathbf v_4(x/4)
    =
    \left(
        1-\frac{3x}{4},\,
        \frac{x}{4},\,
        \frac{x}{4},\,
        \frac{x}{4}
    \right),
    \qquad
    0\leq x\leq1,
    \label{eq:appD-fswap-spectrum}
\end{equation}
which is exactly the fractional SWAP spectrum of
Eq.~\eqref{eq:appA-fswap-spectrum}. Thus the generic four-outcome
upper extremiser is physically attainable throughout its full
parameter range.

\subsubsection{Operator entanglement}

Since
\begin{equation}
    H_2(\mathbf p)
    =
    -\log_2\sum_jp_j^2
    =
    -\log_2(1-E_{\mathrm{op}}),
    \label{eq:appD-H2-Eop}
\end{equation}
fixing $E_{\mathrm{op}}$ is equivalent to fixing $H_2$. Specialising Theorem~2 of
Ref.~\cite{SakaiIwata_2016_Renyi} to fixed order $2$ and target order
$1/2$ gives
\begin{equation}
    H_{1/2}(\mathbf p)
    \leq
    H_{1/2}\!\left(\mathbf v_4(y)\right)
    \label{eq:appD-renyi-upper}
\end{equation}
at fixed $E_{\mathrm{op}}$. Equation~\eqref{eq:appD-gamma-renyi}
therefore makes the fractional SWAP spectrum the global upper
$\gamma$-extremiser over the entire four-component probability
simplex, and Eq.~\eqref{eq:appD-fswap-spectrum} shows that this
extremum is physically attained.

Along fractional SWAP,
\begin{equation}
    E_{\mathrm{op}}
    =
    \frac34x(2-x).
    \label{eq:appD-fswap-Eop}
\end{equation}
Defining
\begin{equation}
    q_E
    =
    \sqrt{1-\frac43E_{\mathrm{op}}},
    \label{eq:appD-qE}
\end{equation}
gives $q_E=1-x$ on $0\leq x\leq1$. Substitution into
$\gamma_{\mathrm{FS}}(x)$ yields
\begin{equation}
    \gamma_{\max}(E_{\mathrm{op}})
    =
    4-3q_E
    +
    3\sqrt{(1-q_E)(1+3q_E)},
    \label{eq:appD-Eop-upper}
\end{equation}
which is Eq.~\eqref{eq:eop-upper-envelope}.

\subsubsection{Schmidt strength}

For fixed Shannon entropy $K_{\mathrm{Sch}}=H_1(\mathbf p)$,
Theorem~1 of Ref.~\cite{SakaiIwata_2016_Shannon}, specialised to
$\alpha=1/2$, makes $\mathbf v_4(y)$ the maximiser of
\begin{equation}
    \|\mathbf p\|_{1/2}
    =
    \left(\sum_j\sqrt{p_j}\right)^2
    =
    \frac{\gamma+1}{2}.
    \label{eq:appD-lhalf-gamma}
\end{equation}
The exact upper boundary is consequently the fractional SWAP
parametric curve
\begin{align}
    K_{\mathrm{FS}}(x)
    &=
    h_2\!\left(\frac{3x}{4}\right)
    +
    \frac{3x}{4}\log_2 3,
    \nonumber\\
    \gamma_{\mathrm{FS}}(x)
    &=
    1+3x+3\sqrt{x(4-3x)},
    \qquad
    0\leq x\leq1.
    \label{eq:appD-K-upper-parametric}
\end{align}
Its entropy parameter is one-to-one. Indeed,
\begin{equation}
    \frac{dK_{\mathrm{FS}}}{dx}
    =
    \frac34
    \log_2
    \left[
        \frac{3(1-3x/4)}{3x/4}
    \right]
    =
    \frac34
    \log_2\!\left(\frac{4-3x}{x}\right),
    \label{eq:appD-KFS-derivative}
\end{equation}
which is positive for $0<x<1$ and vanishes only at the uniform
endpoint $x=1$. Hence Eq.~\eqref{eq:appD-K-upper-parametric}
defines a unique $\gamma_{\max}(K_{\mathrm{Sch}})$ and reproduces
Eq.~\eqref{eq:ksch-upper-envelope-parametric}.

\subsection{Why the physical lower extrema lie on the boundary}
\label{app:entropy-lower-boundary}

We now minimise $\gamma$, equivalently
\begin{equation}
    S(\mathbf p)
    =
    \sum_j\sqrt{p_j},
    \label{eq:appD-S}
\end{equation}
at fixed $E_{\mathrm{op}}$ or $K_{\mathrm{Sch}}$ over the physical
two-qubit spectrum set.

At fixed $H_2$, equivalently fixed purity $\sum_jp_j^2$, an interior
stationary point satisfies
\begin{equation}
    \frac{1}{2\sqrt{p_j}}
    =
    \lambda+2\mu p_j.
    \label{eq:appD-H2-stationarity}
\end{equation}
Writing $x_j=\sqrt{p_j}$ gives
\begin{equation}
    4\mu x_j^3+2\lambda x_j-1=0.
    \label{eq:appD-cubic}
\end{equation}
The derivative of the cubic has at most one positive zero, so the
cubic itself has at most two positive roots. Hence an interior
stationary four-component spectrum has at most two distinct positive
probability values. Up to ordering, the nonuniform possibilities are
\begin{equation}
    (a,b,b,b),
    \qquad
    (a,a,b,b),
    \qquad
    (a,a,a,b),
    \qquad
    a>b>0.
    \label{eq:appD-two-value-spectra}
\end{equation}
The pair-pair case saturates the physical constraint
Eq.~\eqref{eq:physical-schmidt-constraint}. The $3+1$ ordering
$(a,a,a,b)$ violates it because $ab<a^2$. The remaining physical
interior possibility $(a,b,b,b)$ is precisely the generic upper
$H_{1/2}$ extremiser identified above. The boundary minimisation in
Appendix~\ref{app:binary-entropy-allocation} shows that, away from the
deterministic and uniform endpoints, the physical boundary contains a
spectrum with strictly smaller $H_{1/2}$. This interior branch
therefore cannot provide the lower envelope at fixed $H_2$, so the
physical lower $E_{\mathrm{op}}$ extremum lies on the boundary
Eq.~\eqref{eq:appC-boundary-equality} or on a rank-deficient limit.

At fixed Shannon entropy, stationarity gives
\begin{equation}
    \frac{1}{2\sqrt p}
    =
    \lambda+\mu(-\ln p-1).
    \label{eq:appD-H1-stationarity}
\end{equation}
After absorbing constants, define
\begin{equation}
    f(p)
    =
    \frac{1}{2\sqrt p}
    +
    \mu\ln p.
    \label{eq:appD-fp}
\end{equation}
Its derivative,
\begin{equation}
    f'(p)
    =
    \frac{\mu\sqrt p-\frac14}{p^{3/2}},
    \label{eq:appD-fp-derivative}
\end{equation}
has at most one zero, so $f$ is monotone on at most two positive
intervals. Thus $f(p)=\mathrm{constant}$ has at most two positive
solutions, and the same multiplicity classification
Eq.~\eqref{eq:appD-two-value-spectra} applies. The $3+1$ case is
physically excluded, while the $1+3$ case is the generic upper
extremiser at fixed Shannon entropy. The boundary minimisation in
Appendix~\ref{app:binary-entropy-allocation} shows that, away from the
endpoints where the lower and upper extrema coincide, the factorised
boundary contains a spectrum with strictly smaller $H_{1/2}$. Hence
the physical lower $K_{\mathrm{Sch}}$ extremum also lies on the
factorised boundary or its rank-deficient limit.

\subsection{Binary entropy allocation on the physical boundary}
\label{app:binary-entropy-allocation}

Appendix~\ref{app:physical-spectrum-boundary} shows that, up to
permutation, every nontrivial physical-boundary spectrum factorises as
\begin{equation}
    \mathbf p
    =
    \mathbf b\otimes\mathbf c,
    \label{eq:appD-binary-product}
\end{equation}
with two binary probability vectors. R\'enyi additivity gives
\begin{equation}
    H_\alpha(\mathbf p)
    =
    H_\alpha(\mathbf b)
    +
    H_\alpha(\mathbf c).
    \label{eq:appD-renyi-additivity}
\end{equation}
Parameterise a binary spectrum by
\begin{equation}
    \mathbf b_r
    =
    \left(
        \frac{1+r}{2},
        \frac{1-r}{2}
    \right),
    \qquad
    0\leq r\leq1.
    \label{eq:appD-binary-r}
\end{equation}
Then
\begin{align}
    H_{1/2}(\mathbf b_r)
    &=
    \log_2\!\left(1+\sqrt{1-r^2}\right),
    \nonumber\\
    H_2(\mathbf b_r)
    &=
    1-\log_2(1+r^2),
    \nonumber\\
    H_1(\mathbf b_r)
    &=
    h_2\!\left(\frac{1+r}{2}\right).
    \label{eq:appD-binary-entropies}
\end{align}

For the binary alphabet, $H_{1/2}$ is strictly concave as a function
of either $H_2$ or $H_1$. For the $H_2$ relation, write
$s=\sqrt{1-r^2}$. Direct differentiation gives
\begin{equation}
    \frac{d^2H_{1/2}}{dH_2^2}
    =
    (\ln2)
    \frac{
        (s^2-2)(s^2+4s+2)
    }{
        4s^3(1+s)^2
    }
    <
    0
    \label{eq:appD-binary-concavity-H2}
\end{equation}
for $0<r<1$. For the Shannon relation, define
\begin{equation}
    L
    =
    \ln\frac{1+r}{1-r}
    =
    2\,\operatorname{artanh}r.
    \label{eq:appD-L}
\end{equation}
The sign of $d^2H_{1/2}/dH_1^2$ reduces to the positivity of
\begin{equation}
    L(1+2s-s^3)-2r(1+s).
    \label{eq:appD-shannon-concavity-core}
\end{equation}
Since $L>2r$ for $0<r<1$,
\begin{equation}
    L(1+2s-s^3)-2r(1+s)
    >
    2rs(1-s^2)
    >
    0,
    \label{eq:appD-shannon-concavity-positive}
\end{equation}
and therefore
\begin{equation}
    \frac{d^2H_{1/2}}{dH_1^2}<0.
    \label{eq:appD-binary-concavity-H1}
\end{equation}

Let $u$ and $v$ denote the two binary $H_2$ entropies, or the two
binary Shannon entropies, with fixed total $u+v=T$. Strict concavity
implies that the sum of the corresponding two binary $H_{1/2}$ values
is minimised at the boundary of the allowed allocation interval.
Consequently, up to exchange,
\begin{align}
    (u,v)
    &=
    (T,0),
    &
    0\leq T\leq1,
    \nonumber\\
    (u,v)
    &=
    (1,T-1),
    &
    1\leq T\leq2.
    \label{eq:appD-endpoint-allocation}
\end{align}
Thus the lower spectrum first places all available entropy in one
binary factor, and only after that factor becomes maximally mixed does
the second factor acquire entropy. This is the origin of the
controlled-to-pair-pair transition in both lower envelopes.

\subsection{Exact physical operator-entanglement lower envelope}
\label{app:Eop-lower-proof}

For $0\leq E_{\mathrm{op}}\leq1/2$, the endpoint allocation
Eq.~\eqref{eq:appD-endpoint-allocation} gives the rank-two spectrum
\begin{equation}
    \mathbf p
    =
    (t,1-t,0,0),
    \qquad
    \frac12\leq t\leq1.
    \label{eq:appD-Eop-low-spectrum}
\end{equation}
Then
\begin{align}
    E_{\mathrm{op}}
    &=
    2t(1-t),
    \nonumber\\
    \gamma
    &=
    1+4\sqrt{t(1-t)}.
    \label{eq:appD-Eop-low-parametric}
\end{align}
Eliminating $t$ gives
\begin{equation}
    \gamma_{\min}(E_{\mathrm{op}})
    =
    1+2\sqrt{2E_{\mathrm{op}}},
    \qquad
    0\leq E_{\mathrm{op}}\leq\frac12.
    \label{eq:appD-Eop-low}
\end{equation}
These spectra are realised by the controlled family.

For $1/2\leq E_{\mathrm{op}}\leq3/4$, one binary factor is maximally
mixed, giving the pair-pair spectrum
\begin{equation}
    \mathbf p
    =
    \left(
        \frac{t}{2},\,
        \frac{t}{2},\,
        \frac{1-t}{2},\,
        \frac{1-t}{2}
    \right),
    \qquad
    \frac12\leq t\leq1.
    \label{eq:appD-Eop-high-spectrum}
\end{equation}
Then
\begin{align}
    E_{\mathrm{op}}
    &=
    \frac12+t(1-t),
    \nonumber\\
    \gamma
    &=
    3+8\sqrt{t(1-t)}.
    \label{eq:appD-Eop-high-parametric}
\end{align}
Therefore
\begin{equation}
    \gamma_{\min}(E_{\mathrm{op}})
    =
    3+4\sqrt{4E_{\mathrm{op}}-2},
    \qquad
    \frac12\leq E_{\mathrm{op}}\leq\frac34.
    \label{eq:appD-Eop-high}
\end{equation}
Appendix~\ref{app:physical-spectrum-attainability} shows that the
complete pair-pair spectrum locus is physically attainable; CNOT--SWAP
is the canonical representative used in the main text. Combining
Eqs.~\eqref{eq:appD-Eop-low} and \eqref{eq:appD-Eop-high} proves
Eq.~\eqref{eq:eop-lower-envelope}.

\subsection{Exact physical Schmidt-strength lower envelope}
\label{app:K-lower-proof}

The same entropy-allocation result applies directly to
$K_{\mathrm{Sch}}=H_1$. For $0\leq K_{\mathrm{Sch}}\leq1$, the
lower spectrum is
\begin{equation}
    \mathbf p
    =
    (t,1-t,0,0),
    \qquad
    K_{\mathrm{Sch}}=h_2(t),
    \qquad
    \frac12\leq t\leq1.
    \label{eq:appD-K-low-spectrum}
\end{equation}
Using $t=z(K_{\mathrm{Sch}})$ gives
\begin{equation}
    \gamma_{\min}(K_{\mathrm{Sch}})
    =
    1+
    4\sqrt{
        z(K_{\mathrm{Sch}})
        \left[1-z(K_{\mathrm{Sch}})\right]
    }.
    \label{eq:appD-K-low}
\end{equation}

For $1\leq K_{\mathrm{Sch}}\leq2$, one binary factor is uniform:
\begin{equation}
    \mathbf p
    =
    \left(
        \frac{t}{2},\,
        \frac{t}{2},\,
        \frac{1-t}{2},\,
        \frac{1-t}{2}
    \right),
    \qquad
    K_{\mathrm{Sch}}
    =
    1+h_2(t).
    \label{eq:appD-K-high-spectrum}
\end{equation}
Hence $t=z(K_{\mathrm{Sch}}-1)$ and
\begin{equation}
    \gamma_{\min}(K_{\mathrm{Sch}})
    =
    3+
    8\sqrt{
        z(K_{\mathrm{Sch}}-1)
        \left[1-z(K_{\mathrm{Sch}}-1)\right]
    }.
    \label{eq:appD-K-high}
\end{equation}
Equations~\eqref{eq:appD-K-low} and \eqref{eq:appD-K-high} prove
Eq.~\eqref{eq:ksch-lower-envelope}. The switch occurs at the uniform
rank-two spectrum, $K_{\mathrm{Sch}}=1$ and $\gamma=3$.

\subsection{Relation to the generic R\'enyi lower bounds}
\label{app:generic-renyi-lower}

R\'enyi entropy is non-increasing with its order. Hence
\begin{equation}
    H_{1/2}(\mathbf p)
    \geq
    H_2(\mathbf p)
    \label{eq:appD-Hhalf-H2}
\end{equation}
and
\begin{equation}
    H_{1/2}(\mathbf p)
    \geq
    H_1(\mathbf p).
    \label{eq:appD-Hhalf-H1}
\end{equation}
Using Eq.~\eqref{eq:appD-gamma-renyi},
$2^{H_2}=1/(1-E_{\mathrm{op}})$ and
$H_1=K_{\mathrm{Sch}}$, these give
\begin{align}
    \gamma
    &\geq
    \frac{1+E_{\mathrm{op}}}{1-E_{\mathrm{op}}},
    \nonumber\\
    \gamma
    &\geq
    2^{K_{\mathrm{Sch}}+1}-1.
    \label{eq:appD-generic-lower-bounds}
\end{align}
These inequalities hold on the complete probability simplex. Equality
between the distinct R\'enyi orders used here occurs exactly when the
nonzero probabilities are equal, i.e., for a distribution uniform on
its support. For two-qubit unitaries the allowed support
sizes are one, two and four by
Lemma~\ref{lem:physical-schmidt-spectrum}. The operator-entanglement
bound therefore touches the exact physical lower envelope only at
\begin{equation}
    E_{\mathrm{op}}
    =
    0,\qquad
    \frac12,\qquad
    \frac34,
    \label{eq:appD-Eop-Renyi-touch}
\end{equation}
while the Schmidt-strength bound touches it only at
\begin{equation}
    K_{\mathrm{Sch}}
    =
    0,\qquad
    1,\qquad
    2.
    \label{eq:appD-K-Renyi-touch}
\end{equation}
The absent uniform rank-three spectrum would have
$E_{\mathrm{op}}=2/3$ and $K_{\mathrm{Sch}}=\log_2 3$, but is
excluded physically. Thus the generic R\'enyi inequalities remain
valid simple bounds, whereas the piecewise curves derived above are
the sharp physically attainable two-qubit-unitary lower envelopes.
\section{Maximum Concurrence and Perfect Entanglers}
\label{app:cmax-perfect-entanglers}

This appendix proves Theorem~\ref{thm:cmax-envelope} and records the
Cartan geometry underlying Corollary~\ref{cor:perfect-entangler-cost}.
The product-input concurrence formula and perfect-entangler geometry
are inherited from the two-qubit entangling-capacity literature
\cite{Kraus_2001,Chefles_2005,PhysRevA.67.042313}. The general
state-conversion bound for channel quasiprobability extent in
Ref.~\cite{https://doi.org/10.3929/ethz-b-000727956}, together with the pure-state extent
formula there, already implies the lower inequality
$\gamma\geq1+2C_{\max}$. The additional contribution in
Theorem~\ref{thm:cmax-envelope} is to establish global extremal
sharpness at every fixed $C_{\max}$, including the simultaneous sharp
upper boundary and its equality family. For completeness, we also give
a short direct proof of the lower inequality.

\subsection{Concurrence regions in the Cartan tetrahedron}
\label{app:cmax-cartan-regions}

The concurrence formula in Eq.~\eqref{eq:cmax-cartan} is invariant
under $\theta_3\mapsto-\theta_3$, so the geometry may be represented
in the non-negative tetrahedron
\begin{equation}
    0
    \leq
    \theta_3
    \leq
    \theta_2
    \leq
    \theta_1
    \leq
    \frac{\pi}{4}.
    \label{eq:appE-positive-tetrahedron}
\end{equation}
Within this representative tetrahedron, the two planes
\begin{equation}
    \theta_1+\theta_2
    =
    \frac{\pi}{4},
    \qquad
    \theta_2+\theta_3
    =
    \frac{\pi}{4}
    \label{eq:appE-cmax-boundary-planes}
\end{equation}
separate the identity-like and SWAP-like non-perfect-entangler regions
from the perfect-entangler region. The latter is
\begin{equation}
    \theta_1+\theta_2
    \geq
    \frac{\pi}{4},
    \qquad
    \theta_2+\theta_3
    \leq
    \frac{\pi}{4},
    \label{eq:appE-perfect-region}
\end{equation}
where $C_{\max}=1$. Outside this region, the two non-perfect branches
of $C_{\max}$ are those already stated in
Eq.~\eqref{eq:cmax-cartan}.

Several canonical families make the geometry transparent. The
controlled edge meets the perfect-entangler region only at its CNOT
endpoint. The XY/iSWAP family enters it at $\theta=\pi/8$,
corresponding to $\sqrt{\mathrm{iSWAP}}$, while the CNOT--iSWAP edge
lies entirely inside the perfect-entangler region. The CNOT--SWAP
family remains a perfect entangler for
$0\leq\theta\leq\pi/8$ and leaves the region through the SWAP-like
boundary. The fractional SWAP line touches the perfect-entangler
region only at $\theta=\pi/8$, the $\sqrt{\mathrm{SWAP}}$ point.

On the face $\theta_3=0$, the second inequality in
Eq.~\eqref{eq:appE-perfect-region} is automatic. The
perfect-entangler subset of this face is therefore
\begin{equation}
    \theta_1+\theta_2
    \geq
    \frac{\pi}{4},
    \qquad
    0\leq\theta_2\leq\theta_1\leq\frac{\pi}{4}.
    \label{eq:appE-perfect-face}
\end{equation}

\begin{figure}[t]
    \centering
    \includegraphics[width=0.82\linewidth]{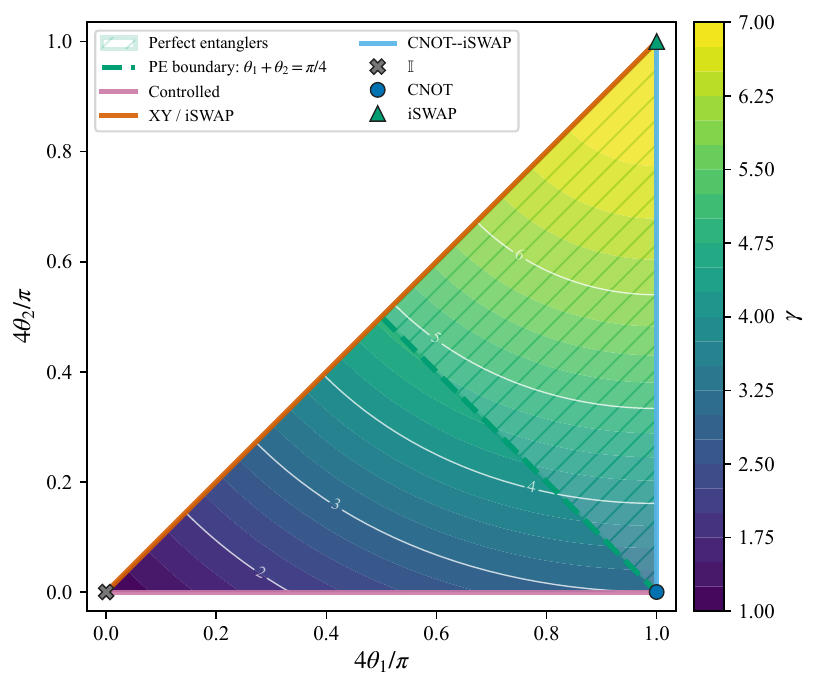}
    \caption{Perfect-entangler geometry on the $\theta_3=0$ face of the
    non-negative Cartan tetrahedron. The perfect-entangler region
    satisfies $\theta_1+\theta_2\geq\pi/4$. CNOT lies at the minimum
    \gls{QPD} extent endpoint $\gamma=3$, while iSWAP lies at the
    uniform-spectrum endpoint $\gamma=7$.}
    \label{fig:perfect-entangler-face}
\end{figure}

The same face geometry is shown in
Figure~\ref{fig:perfect-entangler-face}, together with the
\gls{QPD} extent landscape.

\subsection{A direct nuclear-norm proof of the lower bound}
\label{app:cmax-nuclear-norm}

We now prove the lower bound in Eq.~\eqref{eq:cmax-envelope}. It is
sufficient to work with the Cartan representative
$U_{\mathrm d}$. Indeed, left or right multiplication by local
unitaries preserves the operator-Schmidt singular values and hence
$\gamma$. Local post-unitaries preserve pure-state concurrence, while
local pre-unitaries merely relabel the set of product inputs over which
$C_{\max}$ is maximised. Thus both sides of the desired inequality are
invariant under the local factors in Eq.~\eqref{eq:kak}.

Write
\begin{equation}
    U_{\mathrm d}
    =
    \sum_{j=0}^{3}
    u_j\,\sigma_j\otimes\sigma_j.
    \label{eq:appE-canonical-expansion}
\end{equation}
For an arbitrary normalised product input
$|a\rangle\otimes|b\rangle$, the output is
\begin{equation}
    |\psi\rangle
    =
    \sum_{j=0}^{3}
    u_j
    \bigl(\sigma_j|a\rangle\bigr)
    \otimes
    \bigl(\sigma_j|b\rangle\bigr).
    \label{eq:appE-output-product-expansion}
\end{equation}
Each local factor remains normalised because every $\sigma_j$ is
unitary.

Let $M$ be the $2\times2$ coefficient matrix of $|\psi\rangle$ in a
fixed product basis. Its singular values are the two Schmidt
coefficients $s_1,s_2\geq0$, with
\begin{equation}
    s_1^2+s_2^2=1,
    \label{eq:appE-schmidt-normalisation}
\end{equation}
so its nuclear norm is
\begin{equation}
    \|M\|_*
    =
    s_1+s_2.
    \label{eq:appE-output-nuclear-norm}
\end{equation}
For any expansion of $M$ into rank-one coefficient matrices,
\begin{equation}
    M
    =
    \sum_k
    c_k\,a_k b_k^{T},
    \qquad
    \|a_k\|_2=\|b_k\|_2=1,
    \label{eq:appE-rank-one-expansion}
\end{equation}
the triangle inequality for the nuclear norm gives
\begin{align}
    s_1+s_2
    &=
    \|M\|_*
    \nonumber\\
    &\leq
    \sum_k
    |c_k|\,
    \|a_kb_k^{T}\|_*
    \nonumber\\
    &=
    \sum_k|c_k|,
    \label{eq:appE-nuclear-triangle}
\end{align}
because a rank-one matrix $a_kb_k^T$ has nuclear norm
$\|a_k\|_2\|b_k\|_2=1$. Applying this to
Eq.~\eqref{eq:appE-output-product-expansion} yields
\begin{equation}
    s_1+s_2
    \leq
    \sum_{j=0}^{3}|u_j|.
    \label{eq:appE-schmidt-l1-bound}
\end{equation}
Squaring and using the pure-state concurrence
$C(|\psi\rangle)=2s_1s_2$ gives
\begin{align}
    \left(
        \sum_{j=0}^{3}|u_j|
    \right)^2
    &\geq
    (s_1+s_2)^2
    \nonumber\\
    &=
    1+2s_1s_2
    \nonumber\\
    &=
    1+C(|\psi\rangle).
    \label{eq:appE-l1-concurrence}
\end{align}
Since Eq.~\eqref{eq:gamma-schmidt} is equivalently
\begin{equation}
    \gamma
    =
    2
    \left(
        \sum_{j=0}^{3}|u_j|
    \right)^2
    -1,
    \label{eq:appE-gamma-u-l1}
\end{equation}
we obtain
\begin{equation}
    \gamma
    \geq
    1+2C(|\psi\rangle)
    \label{eq:appE-output-concurrence-bound}
\end{equation}
for every product input. Maximising the right-hand side over all
product inputs therefore proves
\begin{equation}
    \gamma
    \geq
    1+2C_{\max}.
    \label{eq:appE-cmax-lower}
\end{equation}

\subsection{Sharpness of the global envelope}
\label{app:cmax-sharpness}

The lower bound is attained over the full concurrence interval by the
controlled family. For
\begin{equation}
    U_{\mathrm{ctrl}}(\theta)
    =
    \exp\!\left(
        \mathrm{i}\theta X\otimes X
    \right),
    \qquad
    0\leq\theta\leq\frac{\pi}{4},
    \label{eq:appE-controlled-unitary}
\end{equation}
Appendix~\ref{app:family-controlled} gives
\begin{equation}
    \gamma_{\mathrm{ctrl}}
    =
    1+2\sin(2\theta).
    \label{eq:appE-controlled-gamma}
\end{equation}
Acting on $|00\rangle$ gives
\begin{equation}
    U_{\mathrm{ctrl}}(\theta)|00\rangle
    =
    \cos\theta\,|00\rangle
    +
    \mathrm{i}\sin\theta\,|11\rangle,
    \label{eq:appE-controlled-output}
\end{equation}
whose concurrence is $\sin(2\theta)$. Equation~\eqref{eq:cmax-cartan}
shows that this is the maximum product-input concurrence on the
controlled edge, so
\begin{equation}
    C_{\max}^{\mathrm{ctrl}}
    =
    \sin(2\theta),
    \qquad
    \gamma_{\mathrm{ctrl}}
    =
    1+2C_{\max}^{\mathrm{ctrl}}.
    \label{eq:appE-controlled-equality}
\end{equation}
Since $\sin(2\theta)$ spans $[0,1]$, the lower bound is sharp for
every admissible value of $C_{\max}$.

The universal upper bound $\gamma\leq7$ was proved in
Eq.~\eqref{eq:gamma-global-upper}. It is also sharp for every fixed
$C_{\max}$: the iSWAP--SWAP family has $\gamma=7$ throughout, while
Appendix~\ref{app:family-iswap-swap} gives
\begin{equation}
    C_{\max}^{\mathrm{iSWAP-SWAP}}
    =
    \sqrt{1-r^2},
    \qquad
    0\leq r\leq1,
    \label{eq:appE-iswap-swap-cmax}
\end{equation}
which spans $[0,1]$. This completes the proof of
Theorem~\ref{thm:cmax-envelope}.

\subsection{Perfect-entangler consequence}
\label{app:perfect-entangler-proof}

For a perfect entangler, $C_{\max}=1$, so
Eq.~\eqref{eq:appE-cmax-lower} gives $\gamma\geq3$. CNOT lies on the
perfect-entangler boundary and has $\gamma=3$, proving
Corollary~\ref{cor:perfect-entangler-cost}.

The complete interval between the two endpoint costs is also
physically attained within the perfect-entangler class. The
CNOT--iSWAP family remains a perfect entangler throughout and,
from Appendix~\ref{app:family-cnot-iswap},
\begin{equation}
    C_{\max}^{\mathrm{CNOT-iSWAP}}
    =
    1,
    \qquad
    \gamma_{\mathrm{CNOT-iSWAP}}
    =
    3+4\sin(2\theta),
    \qquad
    0\leq\theta\leq\frac{\pi}{4}.
    \label{eq:appE-perfect-interpolation}
\end{equation}
Hence every value in the sharp interval
\begin{equation}
    3
    \leq
    \gamma
    \leq
    7
    \label{eq:appE-perfect-range}
\end{equation}
is realised by a perfect entangler, with CNOT and iSWAP at the two
endpoints.

% ----------------------------------------------------------------
% Author contributions
% ----------------------------------------------------------------
\section*{Author Contributions}

M.H. conceived the project, developed and verified the analytical results,
performed the calculations, prepared the figures, and wrote and revised the
manuscript. A large language model tool was used as an aid for manuscript planning and proofreading, in addition to checking derivations. The author reviewed and verified all AI-assisted material and takes full responsibility for the content of the work.

% ----------------------------------------------------------------
% Bibliography
% ----------------------------------------------------------------
\bibliographystyle{plainnat}
\bibliography{references}

\end{document}